# Childhood Deprivation and Health Inequality in Later Life Across Divergent Life-Course Contexts: Evidence from Estonia, Latvia, and Israel

Nita Handastya

*Department of Economics and Management 'M. Fanno',*

University of Padua, Via Del Santo 33, 35123 Padua, Italy.

**Abstract**

**Background**

Childhood socioeconomic disadvantage is a well-established determinant of health in later life. However, less is known about how early-life deprivation unfolds when individuals experience major institutional transformation and migration in adulthood. Cohorts socialized under Soviet institutions provide a unique setting to examine life-course divergence under systemic change.

**Methods**

This study uses harmonized data on older adult from the Survey of Health, Ageing and Retirement in Europe (SHARE) residing in Estonia, Latvia, and Israel, to examines the association between retrospectively reported childhood deprivation and multiple old-age health outcomes, including: poor self-rated health, chronic diseases burden, functional limitation, depression, and a composite multifrailty indicator. Logistic regression models and predicted probabilities are used to examine whether childhood deprivation predicts late-life health across divergent adult contexts and whether associations vary by linguistic affiliation.

**Results**

Higher levels of childhood deprivation are consistently associated with poorer health in later life across all three countries. Individuals in the highest deprivation quintile exhibit substantially higher odds of adverse health outcomes, including multifrailty, compared with those in the lowest quintile. While the deprivation gradient is present in Estonia, Latvia, and Israel, its magnitude varies across national contexts. Stratification by linguistic affiliation in Estonia and Latvia indicates broadly similar deprivation–health gradients across national-language and Russian-speaking populations, although the precision of estimates differs between groups.

**Conclusion**

Early-life deprivation under Soviet-era conditions is associated with poorer health in older age even after decades of institutional divergence and migration. These findings highlight the long-term persistence of childhood disadvantage and underscore the importance of considering both early-life conditions and adult institutional environments when examining health inequalities in ageing populations exposed to systemic transformation.

## 1. Introduction

The relationship between early-life socioeconomic conditions and old-age health is well established in the life-course epidemiology literature. A large body of research demonstrates that childhood disadvantage can shape health trajectories through multiple mechanism, including cumulative exposure to adverse environments, behavioural pathways, and long-term psychosocial stress (Ben-Shlomo and Kuh, 2002; Barboza Solis et al., 2015; Hughes et al., 2017; Lacey et al., 2020). Individuals who experience material deprivation during childhood are more likely to face poorer health outcomes throughout adulthood, reflecting both the biological embedding of early-life adversity and the accumulation of socioeconomic disadvantage across life course.

Recent empirical studies confirm that retrospective indicators of childhood deprivation remain strongly associated with health outcomes in older populations. Evidence from large European surveys show sthat individuals reporting disadvantaged childhood conditions are more likely to experience poor self-rated health, multimorbidity, functional limitations, and depressive symptoms in later life (Wahrendorf & Blane, 2015; Cheval et al., 2019; Ferraro et al., 2016; Angelini et al., 2016). These findings suggest that early-life socioeconomic circumstances exert long-lasting influences on health trajectories that persist well into old age. However, while a sizeable literature already documents the association, much less is known about whether these relationships persist when individuals experience major institutional transformation or migration across distinct political-economic system.

Most studies examine populations whose life courses unfold within relatively stable institutional environments. Less attention has been given to cohorts whose childhood occurred under one institutional system, but whose adult lives unfolded under significantly different socioeconomic and political conditions. For such cohorts, the relationship between early-life disadvantage and later-life health may evolve under circumstances of institutional disruption, economic transformation, and migration. Individuals socialised under Soviet institutions represent such a cohort. Those born and raised in the Soviet Union were exposed to a broadly shared institutional arrangements, including universal education systems, state-organised employment structures, and publicly provided healthcare. Although living conditions varied across regions and households, these institutions created relatively uniform frameworks of childhood socialisation across Soviet territories.

The dissolution of the Soviet Union in 1991, however, fundamentally altered the institutional environments shaping adult life trajectories. Following this political and economic rupture, individuals who experienced comparable childhood institutional condition entered very different adult contexts. Some remained in newly independent post-Soviet states such as Estonia and Latvia, where rapid market reforms and welfare state restructuring transformed labour markets, social protection systems, and healthcare institutions. Others migrated abroad, including large numbers of Russian-speaking individuals who relocated to Israel during successive migration waves in the late Soviet and early post-Soviet periods. As a result, individuals who shared broadly similar childhood institutional environments were subsequently exposed to significantly different welfare regimes, labour markets, and migration context. This historical event serves as a valuable natural experiment which allows for a life-course study, especially to study whether the long-term consequences of childhood deprivation persist even when later life institutional contexts differ substantially.

This setting raises an important theoretical question for life-course research. If early-disadvantage shapes health through cumulative process and long-term biological embedding, its effects should remain observable regardless of subsequent institutional environments. Conversely, if adult welfare systems, labour markets, and social policies substantially reshape life chances, the influence of childhood deprivation may be mediated or attenuated by the later-life institutional conditions. Examining this cohort therefore provides insight into the persistence of early-life inequality under condition of systemic change.

Within this broader framework, linguistic and migration trajectories introduce an additional dimension of social differentiation. In Estonia and Latvia, Russian-speaking populations occupy distinct social and institutional positions shaped by post-1991 citizenship regimes, language policies, and labour integration process (Cheskin, 2016; Simonyan, 2022). In these contexts, linguistic affiliation reflects more than cultural identity. It is closely linked to patterns of socioeconomic incorporation and access to state institutions in the decades following independence.

At the same time, Russian-speaking migrants who relocated to Israel entered an institutional context fundamentally different from those of the post-Soviet states. Israel operates under a distinct welfare regime, healthcare system, and labour market structure, and had implemented extensive policies aimed at integrating immigrants from the former Soviet Union (Leshem, 2008; Gorodzeisky and Semyonov, 2011); Remennick, 2014). Including this population therefore extends the analytical comparison beyond post-socialist transitions alone. While individuals remaining in Estonia and Latvia experienced institutional transformation within post-Soviet societies, migrants to Israel were incorporated into a non-post-socialist welfare system. This setting provides a contrast to examine the persistence of childhood deprivation across different institutional environments.

This study thus examines the association between childhood deprivation and multiple old-age health indicators among older adults residing in Estonia, Latvia, and Israel. By focusing on individuals whose childhood occurred under Soviet institutional arrangements, but whose adult lives unfolded in different national contexts, the analysis explores whether early-life disadvantage continue to shape health outcomes under divergent institutional trajectories.

Understanding whether early-life socioeconomic disadvantage continues to shape health outcomes across changing institutional environments has important implications for public health policy. If inequalities rooted in childhood persist despite substantial changes in welfare institutions and healthcare systems, this suggests that policies targeting early-life conditions may play a crucial role in reducing long-term health disparities. Examining cohorts who experienced similar childhood conditions but divergent adult contexts therefore provides valuable insight into the persistence of health inequalities over the life course.

This study is thus guided by two questions. First, does childhood deprivation experienced under Soviet-era conditions predict health outcomes in later life across different adult institutional environments? Second, does the magnitude of this relationship vary across national and linguistic contexts representing distinct post-socialist and migration trajectories?

To answer the two questions, this study provides a quasi-experimental perspective on the persistence of early-life inequality. By situating childhood disadvantage within a framework of post-socialist life-course divergence, this study contributes to the literature on ageing and health inequalities in transitional societies. Rather than treating post-Soviet populations as homogenous, this study consider how shared early institutional exposure unfolds across differentiated adult pathways shaped by state transformation and migration.

### 1.1 Soviet Childhood Institutions and Post-1991 Divergence

During the Soviet-era, childhood socialisation occurred within an institutional system system characterised by universal education, state-organised employment structures, and publicly provided healthcare. Although formal access to schooling and primary healthcare was widespread, material shortages, housing constraints, and regional inequalities were persistent in everyday life.

The collapse of the Soviet Union marked a structural break in adult institutional trajectories in several ways. Estonia and Latvia, for instance, implemented rapid market reforms, restructured social insurance systems, and reoriented healthcare financing models. These transformations altered patterns of employment, income security, and old-age securities (Eikemo et al., 2008; Cook, 2013; Aidukaite et al., 2021). These transformations significantly altered patterns of employment, income security, and welfare provision. The dissolution of Soviet institutions eliminated many forms of social protection previously provided by the state, including guaranteed employment.

At the same time, Russian-speaking minorities in the two countries encountered new language requirements and citizenship regimes that shaped their access to labour markets and public institutions. In Estonia and Latvia, not all residents were automatically granted full citizenship after independence, which limited political participation and shaped patterns of socioeconomic incorporation in the newly formed states[1].

In contrast, Russian-speaking migrants who relocated to Israel entered a different institutional framework characterised by integration policies, occupational mobility constraints, and adaptation to a distinct health and welfare system (Leshem, 2008; Gorodzeisky and Semnyonov, 2011; Remennick, 2014). Although many shared Soviet childhood exposures, their adult socioeconomic trajectories were shaped by migration and incorporation into non-post-socialist context.

When taking the two different trajectories, this development created a life-course setting in which early disadvantage occurred under relatively similar institutional conditions, while adult experiences diverged across national and migration context. By examining health outcomes in later life among these groups provides insight into the persistence, and potential mediated external influences, of childhood deprivation effects under systemic transformation.

## 2. Data and methodology

### 2.1 Sample and data

This study uses data from the Survey of Health, Ageing and Retirement in Europe (SHARE), a multi-disciplinary, cross-national European study of individuals aged 50 and older[2]. SHARE collects harmonised information on health, socioeconomic status, and social networks across European countries and Israel. Since its launch in 2004, the survey has been conducted biennially and provides detailed information on ageing population in diverse institutional contexts. Of relevance for this study is the SHARELIFE module, which collects retrospective life-history information, including indicators of childhood socioeconomic conditions. These retrospective measures enable the examination of long-term association between early-life deprivation and later-life health outcomes.

The analysis focuses on respondents residing in Estonia, Latvia, and Israel. Estonia and Latvia represent post-socialist societies that experienced rapid institutional transformation following the dissolution of the Soviet Union, while Israel serves as a migration-based comparison case. A substantial share of Israel's older population migrated from the former Soviet Union during two major waves of Aliyah, first in the 1970s and later in the early 1990s following the collapse of the Soviet Union. As a result, Israel contains a sizeable population of older Russian-speaking migrants who share early-life institutional experiences with individuals who remained in post-Soviet states. Including Israel, therefore, allows comparison between individuals who experienced similar childhood conditions under Soviet institutions, but whose adult lives unfolded in different institutional environments.

SHARE survey was conducted in multiple languages in the three countries. Survey in Latvia and Estonia respectively were conducted in Latvian/Russian and Estonian/Russian. The Israeli survey was conducted in three languages: Hebrew, Russian, and Arabic. Arabic-speaking respondents are excluded from the analysis because the focus of this study is on individuals whose childhood socialisation occurred within Soviet institutional contexts. Restricting the Israeli sample in this way improved comparability with respondents in Estonia and Latvia, and ensures that the analysis focuses on populations with broadly similar early-life institutional exposure. Lithuania, although a part of the Baltic region, is excluded from this study due to the lack of Russian language module.

---

[1] It is important to note that for Estonia and Latvia, the event that unfolded in 1991 was considered a restoration of independence. Thus, automatic citizenship was only given to those whose ascendants were already a citizen prior to the annexation by the Soviet Union in 1940.

[2] See Börsch-Supan et al. (2013) and Weber (2018) for more details about SHARE data.

The empirical analysis in this study uses data from SHARELIFE (wave 7), together with wave 8 and wave 9 for the SHARE survey (SHARE-ERIC, 2024a; 2024b; 2024c). SHARELIFE provides retrospective information on respondents' childhood circumstances. Wave 7, 8, and 9 contains the most recent information on health outcomes and socioeconomic characteristics in later-life. Estonia Joined the SHARE survey in wave 4 (2010/2011), Latvia in wave 7 (2017), and Israel participated since wave 1 (2005/2006) with the exception of wave 3 and 4.

The sample is restricted to respondents aged 50 or older at the time of the wave 7 interview who participated in at least one of three waves used in the analysis. Observations from these waves are pooled to increase statistical power and to incorporate the most recent available information on health outcomes. Because some respondents appear in multiple waves, standard errors are clustered at the individual level to account for repeated observations.

### 2.1.1 Health outcomes

Four indicators are used to represent late-life health outcomes: (1) poor self-rated health, (2) chronic diseases burden, (3) functional limitation, and (4) depression. Health information is drawn from Waves 7 through 9.

Poor self-rated health is constructed as a binary variable based on the question: "Would you say your health is… Excellent, Very good, Good, Fair, or Poor?" The indicator equals 1 if the respondent reports "Fair" or "Poor," and 0 otherwise.

Chronic disease burden is based on whether a doctor has ever told the respondent that they had any of 21 listed conditions[3] for the respondent to note. A binary indicator is created taking the value 1 if the respondent reports two or more chronic conditions and 0 otherwise.

Functional limitation is derived from two sets of questions on ADL (Activities of Daily Living) and IADL (Instrumental Activities of Daily Living). Respondents are asked whether they have difficulty performing a list of everyday activities expected to last more than three months. The enumerator then read out a list of 10 activities[4] which accounts for the ADL, and another list of 15 activities[5] on IADL for the respondent to note. A binary indicator equals 1 if the respondent reports difficulty in at least one activity and 0 otherwise.

[3] List of conditions presented to the respondent: a heart attack including myocardial infarction or coronary thrombosis or any other heart problem including congestive heart failure; high blood pressure or hypertension; high blood cholesterol; a stroke or cerebral vascular disease; diabetes or high blood sugar; chronic lung disease such as chronic bronchitis or emphysema; cancer or malignant tumor, including leukaemia or lymphoma, but excluding minor skin cancers; stomach or duodenal ulcer, peptic ulcer; Parkinson's disease; cataracts; hip fracture; other fractures; Alzheimer's disease, dementia, organic brain syndrome, senility or any other serious memory impairment; other affective or emotional disorders, including anxiety, nervous or psychiatric problems; rheumatoid arthritis; chronic kidney disease.

[4] List of activities read to the respondent: walking 100 metres; sitting for about two hours; getting up from a chair after sitting for long periods; climbing several flights of stairs without resting; climbing one flight of stairs without resting; stooping, kneeling, or crouching; reaching or extending your arms above shoulder level; pulling or pushing large objects like a living room chair; lifting or carrying weights over 10pounds/5kilos, like a heavy bag of groceries; picking up a small coin from a table.

[5] List of activities read to the respondent: dressing, including putting on shoes and socks; walking across a room; bathing or showering; eating, such as cutting up your food: getting in or out of bed; using a toilet, including getting up or down; using a map to figure out how to get around in a strange place; preparing a hot meal; shopping for groceries; making telephone calls; taking medications; doing work around the house of garden; managing money, such as paying bills and keeping track of expenses; leaving the house independently and accessing transportation services; doing personal laundry.

Depression is measured using the EURO-D scale (range 0–12), where higher values indicate more depressive symptoms. It is based on the EURO-D scale that was developed by European Commission[6]. Following established practice, a binary indicator of depression "caseness" is created, with scores of 4 or higher coded as 1 while 3 or lower as otherwise. This measure is available for Waves 8 and 9 only, resulting in a smaller sample for models including depression.

**Multifrailties**

The second set of indicators aim to capture multifrailties experienced by the respondent. Four indicators are constructed based on the previous four health outcomes, namely: 1) poor self-rated health AND functional limitation, 2) poor self-rated health AND chronic diseases burden, 3) chronic diseases burden AND functional limitation, 4) poor self-rated health AND functional limitation AND chronic diseases burden.

This composite measure is designed to capture overlapping subjective and objective dimensions of health vulnerability. Self-rated health may reflect both medical conditions and culturally shaped perception of health, while functional limitation and chronic diseases burden provide more objective measures of physical functioning. For example, a study from China by Wu et al. (2013) found that obesity is unrelated to a poor self-rated health. They elaborated that it is caused by a view in Chinese culture in which obesity is not a signal of unhealthiness but rather a sign of wealth. By combining these indicators, the analysis assesses whether childhood deprivation is associated with multifrailty rather than a singular health outcome.

### 2.1.2 Childhood deprivations indicator

This study utilises a childhood deprivation score constructed from available retrospective information that was asked to respondents on wave 7. Four dimensions are considered: living condition, health related deprivation, financial hardship, and emotional deprivation. Each dimension is constructed from multiple binary indicators and are given equal weight. A total childhood deprivation score is calculated as the equally weighted sum of these dimensions, scaled between 0 and 1. Higher values indicate greater deprivation. For the analysis, the total score is divided into country-specific quintiles.

**Living conditions**

The first dimension is the living condition deprivation score. It is constructed based on three indicators: housing deprivation, overcrowding, and availability of books at home. The three indicators are based on survey questions from the childhood circumstances module of SHARELIFE wave 7.

Housing deprivation is constructed based a survey question in which the respondent is asked to recall the accommodation in which they lived in when they were 10 years old, and whether they have 5 amenities[7] listed by the enumerator. A binary variable is generated in which a value 1 will be assigned if the respondent noted at least 3 out of 5 amenities, and 0 otherwise.

The overcrowding indicator is constructed based on a survey question in which the respondent is asked about how many rooms and housemates there were in the accommodation. I then calculated the number of persons per room, and set a cut-off point of 2 people per room as an indicator of overcrowding. The resulted variable is a binary with 1 indicating overcrowding.

Lastly, the availability of the books is based on a question worded "Approximately how many books were there in the place you lived in when you were 10? Do not count magazines, newspapers, or your schoolbooks". There are 5 possible answers: "None or very few (0-10 books)", "Enough to fill one shelf (11-

---

[6] Further detail regarding this measure can be found on Prince et al. (1999). Further discussion on its application specifically to SHARE data can be found on Maskileyson et al. (2021)

[7] List of housing amenities presented to the respondent: fixed bath; cold running water supply; hot running water supply; inside toilet; central heating.

25 books)", "Enough to fill one bookcase (26-100 books)", "Enough to fill two bookcases (101-200 books)"," Enough to fill two or more bookcases (more than 200 books)". The answer is then dichotomised with the value of 1 if the respondent chose "None or very few (0-10 books)" or "Enough to fill one shelf (11-25 books)", and 0 otherwise.

**Health-related deprivation**

The second dimension is the health deprivation score. There are five indicators used in the creation of this score: vaccination, doctor's visit, childhood illness, dental care, and subjective health.

The vaccination is based on a question worded "During your childhood, that is, before you turned 16, have you received any vaccinations?". The response is binary with 1 representing a "Yes" and 0 a "No", but it is recoded to note non-vaccination instead.

The doctor's visit is based on a question worded "Not including dental care, have you ever needed to see a doctor but you did not because you could not afford it?". The response is binary with 1 representing a "Yes" and 0 a "No".

The childhood illness indicator is based on two questions worded similarly: "Did you have any of the diseases on this card during your childhood (that is, from when you were born up to and including age 15)?". Collectively, there are 20 types of childhood illnesses[8] listed. I created a cut-off point of 2 cumulative conditions. The resulted indicator holds a value of 1 if the respondent had at least 2 childhood illnesses and 0 otherwise.

The dental care indicator is constructed by two questions worded respectively "Did you start going regularly to the dentist during your childhood (that is, before you turned 16)?" and "Please specify the periods in which you did not go to a dentist for checkups or dental care regularly." The former yielded a binary response of yes and no, while the latter listed multiple life periods. For the creation of this indicator, I only consider the "When I was 0-15 years old". The second question is only used when the first demonstrated a missing value. The indicator is further transformed to represent absence of dental care as a child.

The subjective health indicator is based on a question worded "Would you say that your health during your childhood was in general excellent, very good, good, fair, or poor?" with five responses: "Excellent", "Very good", "Good", "Fair", "Poor". Based on the answer, I created a binary indicator named "Poor subjective health" that would take a value of 1 if the responded answered either "Fair" or "Poor".

**Financial hardship**

The third dimension is the financial deprivation score. It is based only on one variable worded "Now think about your family when you were growing up, from birth to age 16. Would you say your family during that time was well off financially, about average, or poor". There are five possible responses: "Pretty well off financially", "About average", "Poor", "It varied", "Did not live with family". The responses are dichotomised to a binary indicator with the value 0 if the respondent answered, "Pretty well off financially" or "About average" and 1 otherwise.

[8] Listed childhood illnesses: Infectious disease (e.g. measles, rubella, chickenpox, mumps, tuberculosis, diphtheria, scarlet fever); Polio; Asthma; Respiratory problems other than asthma; Allergies (other than asthma); Severe diarrhoea; Meningitis/encephalitis; Chronic ear problems; Speech impairment; Difficulty seeing even with eyeglasses; Severe headaches or migraines; Epilepsy, fits or seizures; Emotional, nervous, or psychiatric problem; Broken bones, fractures; Appendicitis; Childhood diabetes or high blood sugar; Heart trouble; Leukaemia or lymphoma; Cancer or malignant tumour (excluding minor skin cancers); Rickets, osteomalacia, rachitis.

**Emotional deprivation**

The fourth dimension is the emotional wellbeing deprivation score. It is based primarily on the respondent's relationship with their parents. Three sets of questions are asked of each parent, yielding 6 responses. The first question is worded "Which of the people on this card did you live with at this accommodation when you were 10?" with 9 possible answers. However, the response of interests is only "Biological mother" and "Biological father". The response is 1 if the answer was no, and 0 otherwise. It indicated absence of biological parents. The second question is worded "How much did your mother father understand your problems and worries?" with four possible responses: "A lot" "Some" "A little" "Not at all". The response is dichotomised to 1 if the response is "A little" or "Not at all" and 0 otherwise to indicate lack of feeling of being understood. The third question concerns physical harm. The question is worded "How often push, grab, shove, throw something at you, slap or hit you?" with four possible responses: "Often" "Sometimes" "Rarely" "Never". The response is dichotomised to 1 if the response is "Often" or "Sometimes" and 0 otherwise.

The total childhood deprivation score is constructed following the multidimensional deprivation framework proposed by Alkire and Foster (2011). Each deprivation indicator is assigned equal weight, and the indicators are aggregated to form a composite measure of childhood deprivation. The resulting index is normalised to range between 0 and 1, where higher values indicate greater levels of childhood deprivation.

To facilitate interpretation and allow for nonlinear association, the continuous index is subsequently divided into country-specific quintiles. Individuals in the first quintile represent those who experienced the lowest level of childhood deprivation, while those in the fifth quintile correspond to respondents with the highest level of deprivation. Using country-level quintiles accounts for differences in the distribution of childhood conditions across the three countries while preserving the relative deprivation ranking within each national contexts.

**2.1.3 Linguistic affiliation**

Linguistic affiliation is proxied by the language in which the SHARE interview was conducted. Respondents interviewed in Russian are classified as Russian speakers, while those interviewed in the national language (Estonian or Latvian or Hebrew) are classified as national-language speakers.

Using interview language provides a consistent indicator of linguistic affiliation across the countries included in the analysis. Although interview language does not fully capture ethnicity, citizenship status, or migration history, it serves as a practical proxy for everyday language use and linguistic incorporation. In Estonia and Latvia in particular, language use is closely associated with patterns of socioeconmic integration and access to state institutions following independence.

**2.1.4 Control variables**

The study used three set of covariates that acted as the control. The three sets are demographic, household characteristics, and current circumstances.

The demographic set contains gender, age, education, country of residence, and employment status. I introduced additional squared age and cubic age to capture a possible non-linear relationship between age and the dependent variables.

Household characteristics contains an indicator of cohabitation with a partner, urban or rural dwelling, and household size. I used three indicators regarding income: equivalised household income, inverse hyperbolic sine transformation of the household income, and income quartile of the household.

Current circumstances contain an indicator of the ability to make ends meet and place of birth.

**2.2 Empirical approach**

To examine the association between childhood deprivation and late-life health, I estimate logistic regression models of the following form:

$$logit(Y_{it}) = \alpha + \beta_1 CD_i + \gamma X_{it} + \delta_c + \tau_t + \epsilon_{it}$$

*where*

- $Y_{it}$ denotes one of the four late-life health outcomes for individual *i* at time *t*:
  (1) poor self-rated health,
  (2) chronic disease burden,
  (3) functional limitation, or
  (4) depression.
- $\mathrm{CD}_i$ represents quintiles of the total childhood deprivation score (reference category: first quintile).
- $\mathbf{X}_{it}$ is a vector of individual and household-level covariates, including demographic characteristics, socioeconomic indicators, and current living conditions.
- $\delta_c$ denotes country fixed effects (Estonia as reference in pooled models).
- $\tau_t$ represents wave fixed effects.
- $\varepsilon_{it}$ is the error term.

The primary specification pools Estonia, Latvia, and Israel in order to estimate the overall gradient between childhood deprivation and late-life health while adjusting for country-level differences. Estonia serves as the baseline, allowing direct comparison with Latvia and Israel. To assess whether the deprivation-health gradient differs across Baltic contexts, separate models are estimated for Estonia and Latvia using the same covariate structure but excluding country indicators. Israel is not included in these country-stratified models, as its role in the analysis is to serve as a post-socialist migrant comparison group rather than a symmetric Baltic case. Given the central focus on linguistic affiliation, additional models are estimated separately for Russian-speaking and national-language-speaking respondents in Estonia and Latvia. Linguistic affiliation is proxied by the language of interview. These stratified models allow comparison of the magnitude of the deprivation gradient across linguistic groups. Finally, interaction models between childhood deprivation quintiles and linguistic affiliation are estimated in pooled specifications to formally test whether the association between early-life deprivation and late-life health differs between Russian-speaking and national-language-speaking respondents.

All models adjust for gender; age (linear and quadratic terms); education; employment status; cohabitation status; urban residence; household size; income quartile; difficulty in making ends meet; place of birth; and wave fixed effects.

Odds ratios (OR) are reported for ease of interpretation. The study is carried out with Stata 18.

### 2.3. Limitations

Several limitations should be considered when interpreting the findings of this study. First, the childhood indicators are based on recall method and may therefore be subject to recall bias. However, previous validation studies suggest that retrospective information collected in the SHARELIFE module provides reasonably reliable measures of childhood circumstances (Havari and Mazzona, 2015).

Second, the sample only includes those who survived to the age of 50+. This may lead to selective mortality bias, whereby individuals exposed to severe childhood deprivation are less likely to survive into older age. As a result, the estimated association between early-life deprivation and later-life health may be attenuated (Case and Paxson, 2010; Barboza Solis et al, 2015).

Third, Adult socioeconomic conditions may partly mediate the relationship between childhood circumstances and health in later life; therefore, controlling for these variables primarily captures the direct association rather than the total life-course effect (Ben-Shlomo & Kuh, 2002).

Fourth, the study also does not distinguish between Russian speakers who migrated to Estonia and Latvia during the Soviet period and those whose families had settled in the region prior to the independence. These group may differ in theis socioeconomic integration and identity formation, which could influence health outcomes in later life (Cheskin, 2021).

## 3. Results

### 3.1 Sample selection

The analytic sample includes respondents aged 50 and above who participated in SHARE waves 7 to 9 and provided complete information on childhood conditions and health outcomes. After excluding observations with missing values, the final sample consists of 10,079 observations corresponding to 4,771 unique individuals.

Table 1 presents descriptive statistics for the pooled sample and by country. Estonia accounts for approximately 64 % of observations, Latvia for 21%, and Israel for 15%. The average age of respondents is 70.2 years, and roughly two-thirds of the sample are female.

The childhood deprivation index ranges from 0 to 1, with higher values indicating greater deprivation. Mean levels are relatively similar in Estonia (0.27) and Latvia (0.24), suggesting broadly comparable early-life conditions among the current elderly cohorts. Israel displays a lower average score (0.21). However, the number is largely similar.

Health indicators reveal substantial variation across countries. Overall, 63.9% of respondents report poor self-rated health. This proportion is highest in Estonia (70.1 %), followed by Latvia (64.4 %), and considerably lower in Israel (35.8 %). The prevalence of chronic diseases burden (two or more diagnoses) averages roughly 52.3% across countries with the lowest figure exhibited by Israel (47.6%) while Estonia stands at 54%. Functional limitations (more than one ADL/IADL difficulty) are more common in Estonia (25.7 %) than in Latvia (19.8 %). Depressive caseness, measured using the EURO-D caseness and only available for waves 8 and 9, is also more prevalent in Estonia (33.8%) than in Latvia (26.0%).

In terms of socioeconomic characteristics, nearly half of respondents have completed secondary education, and approximately one-third have tertiary education. In both Estonia and Latvia, more than half of the respondents attained secondary education, while Israel shows the highest proportion of tertiary education attainment at 47.3%. Around 58 % live with a partner, and the average household size is below two persons, reflecting the predominance of smaller households in older age. The figure from Israel notes that 73.9% of the respondents live with their partner. This might reflect the uneven life expectation between gender. Latvia demonstrates the highest gap in European Union where women's figure is around 10 years above men, followed by 8 years in Estonia and 4 years for Israel. Majority of the respondents live in an urban area, but the number is unusually high for Israel with 97.3% positive response. A substantial share of respondents report difficulty making ends meet, particularly in Latvia, where more than 60 % indicate financial strain.

Majority of the respondents are national language speaker, with an average of 15% across the three countries who speak another language (in the context of this study, Russian language). In Estonia and Latvia, respectively 18.5% and 17.4% of the respondent were born elsewhere as opposed to 55.7% in Israel. However, the citizenship indicator provides an interesting insight. Although more than half of Israeli respondents were born abroad, 99.8% are citizens. Meanwhile, Estonia and Latvia which exhibit lower extranational birth only around 84-88% possess citizenship status.

*Table 1 Descriptive statistics*

| | All (N=10,079) 100% | | Estonia (N=6,487) 64.36% | | Latvia (N=2,107) 20.90% | | Israel (N=1,485) 14.73% | |
|---|---|---|---|---|---|---|---|---|
| | Mean | SD | Mean | SD | Mean | SD | Mean | SD |
| Total childhood deprivation score | 0.2566 | | 0.2708 | | 0.2437 | | 0.2129 | |
| | | | | | | | | |
| Poor self-rated health | 0.6385 | | 0.7012 | | 0.6435 | | 0.3575 | |
| Chronic diseases burden | 0.5225 | | 0.5398 | | 0.5026 | | 0.4754 | |
| Functional limitation | 0.2328 | | 0.2565 | | 0.1979 | | 0.1791 | |
| Depression (N=5,318) | 0.3213 | | 0.3380 | | 0.2600 | | 0.3259 | |
| | | | | | | | | |
| Is a female | 0.6644 | | 0.6806 | | 0.6563 | | 0.6047 | |
| Age | 70.1658 | 9.6342 | 70.94 | 9.7101 | 67.3910 | 9.6331 | 70.6821 | 8.5255 |
| Education | | | | | | | | |
| None | 0.0043 | | 0.0011 | | 0.0030 | | 0.0195 | |
| Primary | 0.1674 | | 0.1854 | | 0.0982 | | 0.1865 | |
| Secondary | 0.4965 | | 0.5078 | | 0.5852 | | 0.3212 | |
| Tertiary | 0.3319 | | 0.3057 | | 0.3132 | | 0.4727 | |
| Is living with a partner | 0.5841 | | 0.5554 | | 0.5633 | | 0.7387 | |
| Household size | 1.9056 | | 1.8445 | 0.8363 | 1.9729 | 0.9753 | 2.0740 | 0.9006 |
| Lives in an urban area | 0.7608 | | 0.7492 | | 0.6473 | | 0.9723 | |
| Income quartile | | | | | | | | |
| First | 0.2288 | | 0.2209 | | 0.2682 | | 0.2074 | |
| Second | 0.2531 | | 0.2577 | | 0.2444 | | 0.2451 | |
| Third | 0.2643 | | 0.2638 | | 0.2340 | | 0.3098 | |
| Fourth | 0.2538 | | 0.2576 | | 0.2534 | | 0.2377 | |
| Employment status | | | | | | | | |
| Retired | 0.6139 | | 0.6405 | | 0.5956 | | 0.5239 | |
| Employed or self-employed | 0.3051 | | 0.3018 | | 0.3090 | | 0.3138 | |
| Others | 0.0810 | | 0.0577 | | 0.0954 | | 0.1623 | |
| Having difficulties to make ends meet | 0.4572 | | 0.4314 | | 0.6468 | | 0.3010 | |
| Language of the interview: national | 0.8435 | | 0.8247 | | 0.8879 | | 0.8626 | |
| | | | | | | | | |
| Respondent was born elsewhere | 0.2375 | | 0.1851 | | 0.1737 | | 0.5569 | |
| Respondent is a citizen | 0.8911 | | 0.8831 | | 0.8410 | | 0.9973 | |

### 3.3 Regression outcomes

This section presents results from multivariable regression models assessing whether childhood deprivation remains associated with health outcomes in later life after adjusting for demographic, socioeconomic, and household characteristics. The pooled specification includes respondents from Estonia, Latvia, and Israel and controls for age, gender, education, employment status, household composition, income quartile, migration background, and wave fixed effects. Standard errors are clustered at the individual level. Estonia serves as the reference category for country comparisons. The main findings are presented through figures showing predicted probabilities and odds ratios across childhood deprivation quintiles, while the full regression tables are reported in the appendix.

Figure 1 presents predicted probabilities of four late-life health outcomes—poor self-rated health, chronic disease burden, functional limitation, and depression—across childhood deprivation quintiles for Estonia, Latvia, and Israel, using the lowest deprivation quintile (Q1) as the reference category. Across most outcomes, a general upward pattern is visible. Individuals who experienced higher levels of childhood deprivation tend to display higher predicted probabilities of adverse health outcomes in older age. The increase is most pronounced when comparing the lowest deprivation quintile (Q1) with the highest (Q5), although the magnitude and smoothness of the gradient vary across outcomes and countries.

The pattern is particularly clear for chronic disease burden and functional limitation. For these outcomes, predicted probabilities generally increase as childhood deprivation rises, with the highest quintile typically displaying the largest values across countries. Functional limitation and depression show especially visible increases between the lower and upper deprivation quintiles, particularly in Estonia and Israel. By contrast, the relationship between childhood deprivation and poor self-rated health appears somewhat flatter for Estonia and Latvia, where predicted probabilities are already relatively high at the lowest deprivation levels, while Israel shows a clearer upward trend across quintiles.

Cross-country differences are also visible in the overall levels of predicted probabilities. Estonia and Latvia display higher probabilities of poor self-rated health than Israel across all deprivation quintiles. For chronic disease burden and depression, predicted probabilities are more similar across countries, although Estonia tends to show slightly higher values at the upper end of the deprivation distribution. Overall, the figure indicates that childhood deprivation is associated with multiple dimensions of health in later life while also revealing differences in the level and shape of the relationship across health outcomes and national contexts. Taken together, these descriptive patterns provide initial evidence of a deprivation gradient in later-life health, which is examined more formally in the regression results that follow.

*Figure 1 Gradient of old-age health outcomes across childhood deprivation quintiles*

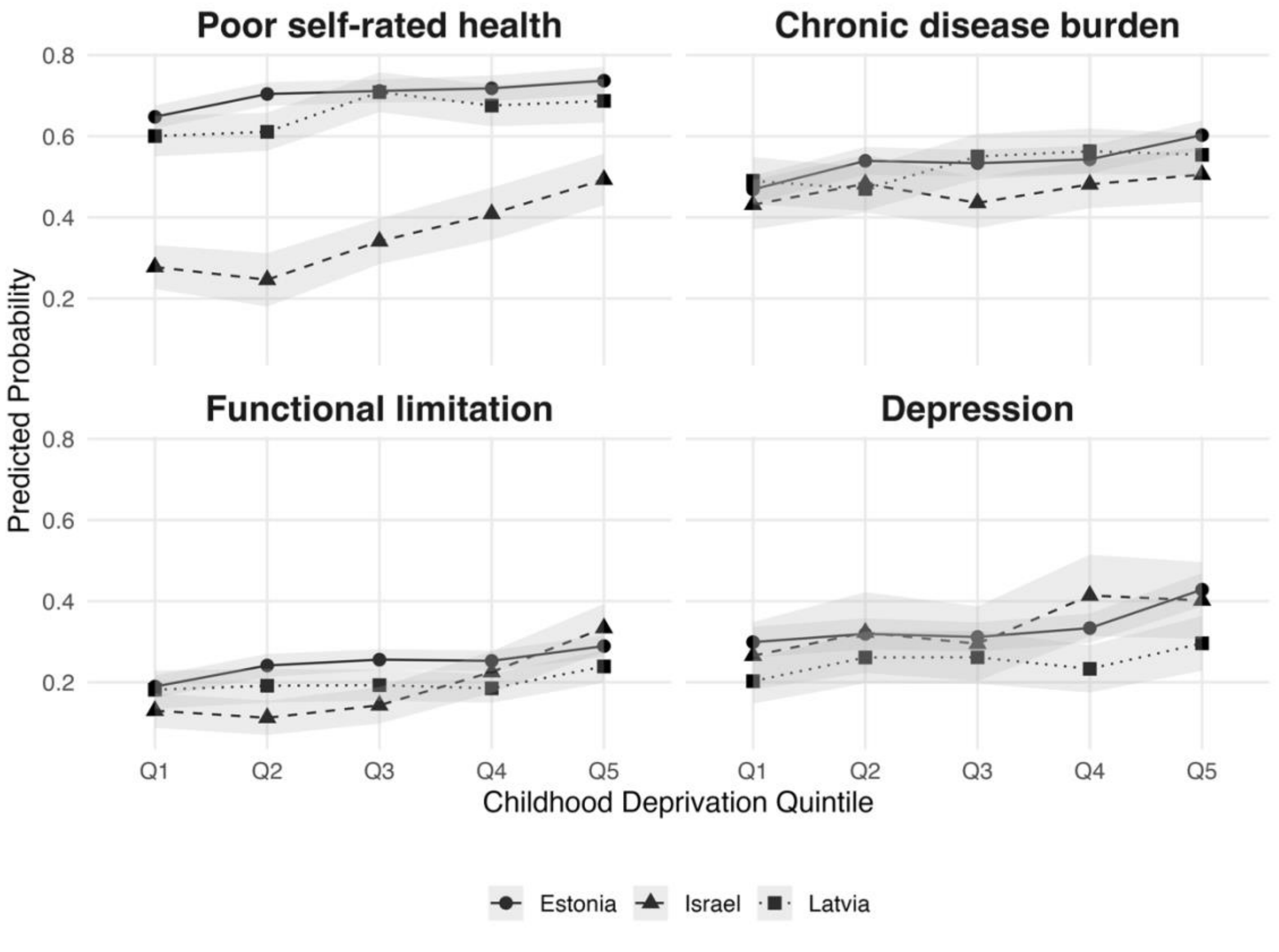

The next figure presents predicted probabilities of multifrailty across childhood deprivation quintiles for Estonia, Latvia, and Israel. In this context, multifrailty is defined as the joint presence of poor self-rated health, chronic disease burden, and functional limitation. Overall, a generally increasing pattern is visible across the deprivation distribution. Individuals in higher childhood deprivation quintiles tend to exhibit higher predicted probabilities of experiencing multifrailty in later life compared with those in the lowest quintile.

The deprivation gradient is observable in all three countries, although the shape and steepness differ somewhat across contexts. Estonia shows a gradual increase in predicted probabilities from the lowest to the highest deprivation quintile, with a small fluctuation around the fourth quintile before rising again at the highest level. Israel displays a more pronounced gradient beginning in the middle of the distribution, with predicted probabilities increasing sharply between the third and fifth quintiles. Latvia exhibits a more moderate pattern, with relatively small differences across the lower quintiles and a clearer increase toward the upper end of the deprivation distribution.

Despite these differences in slope, the highest deprivation quintile is associated with the largest predicted probability of multifrailty in all three countries. Cross-country differences are also visible in the overall levels of predicted probability. Estonia and Israel show relatively higher probabilities at the upper end of the deprivation distribution, while Latvia displays a more gradual increase across quintiles. Taken together, these results suggest that childhood deprivation is associated not only with individual health outcomes but also with the accumulation of multiple health vulnerabilities in later life.

*Figure 2 Gradient of composite multifrailty in old- health outcomes across childhood deprivation quintiles*

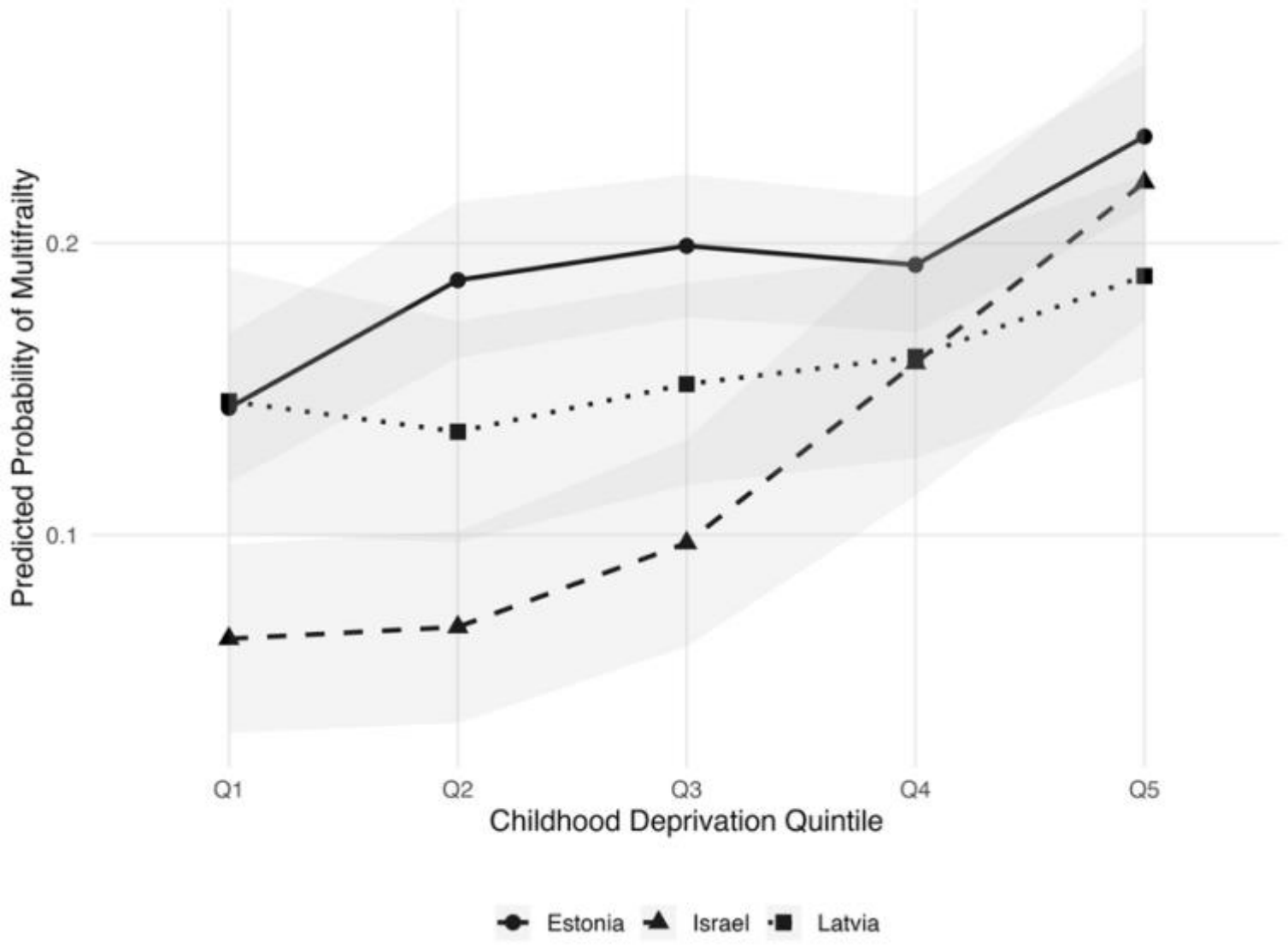

The next part of the analysis examines whether the association between childhood deprivation and later-life health differs by linguistic affiliation in Estonia and Latvia. Figure 3 presents estimated odds ratios for four old-age health outcomes—poor self-rated health, chronic disease burden, functional limitation, and depression—stratified by interview language. Respondents interviewed in Russian are classified as Russian speakers, while those interviewed in the national language (Estonian or Latvian) are classified as national-language speakers. The lowest childhood deprivation quintile (Q1) serves as the reference category. Points represent estimated odds ratios for each quintile relative to Q1, and horizontal bars indicate confidence intervals.

Overall, the figure shows that higher levels of childhood deprivation are generally associated with higher odds of adverse health outcomes in later life for both linguistic groups. Across the four outcomes, most estimated odds ratios exceed one, particularly in the higher deprivation quintiles, indicating increased risks relative to individuals who experienced the lowest level of childhood deprivation.

For poor self-rated health, chronic disease burden, and functional limitation, both linguistic groups display higher odds ratios in the upper deprivation quintiles (Q3–Q5). Among national-language speakers, the increase in odds ratios tends to follow a more gradual pattern across the deprivation distribution. For Russian speakers, the estimates show somewhat greater variation across quintiles. In several cases, the confidence intervals for Russian speakers are wider, reflecting greater statistical uncertainty in these estimates.

A similar pattern is observed for depression. Odds ratios increase across deprivation quintiles for both linguistic groups, particularly in the higher part of the distribution. Although the magnitude of the estimates varies across outcomes and groups, the overall pattern suggests that individuals who experienced greater childhood deprivation face elevated risks of adverse health outcomes in later life regardless of linguistic affiliation. Differences in the size and precision of the estimates indicate that the strength of the association may vary somewhat across linguistic groups; however, the overall deprivation gradient remains visible in both populations. This suggests that the long-term health consequences of childhood disadvantage operate in broadly similar ways across national-language and Russian-speaking groups in the Baltic context.

*Figure 3 Correlation between childhood deprivation quantiles and old-age health by linguistic group in Latvia and Estonia*

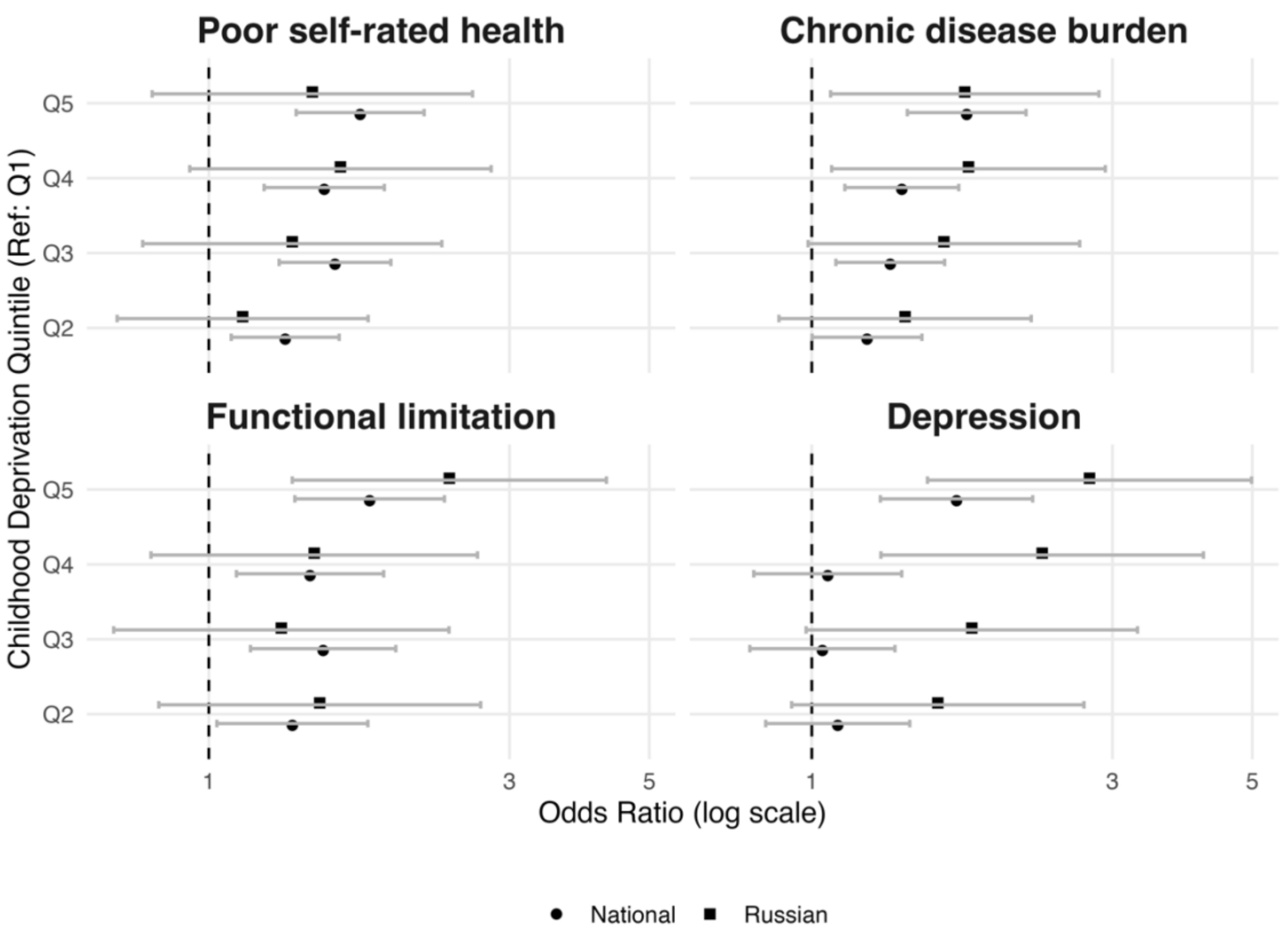

Figure 4 presents the estimated odds ratios for multifrailty across childhood deprivation quintiles using the same specification as in Figure 4. Multifrailty is defined as the joint presence of poor self-rated health, chronic disease burden, and functional limitation. The lowest deprivation quintile (Q1) serves as the reference category.

Overall, the figure indicates that higher levels of childhood deprivation are associated with increased odds of experiencing multifrailty in later life. The increase is most evident at the upper end of the deprivation distribution, where individuals in the highest quintile (Q5) display substantially higher odds relative to those in the lowest deprivation group.

The pattern differs somewhat between linguistic groups. Among national-language speakers, the estimated odds ratios increase more gradually across the deprivation distribution. For Russian speakers, the point estimates appear somewhat larger in the highest deprivation quintile, suggesting a steeper increase toward the upper end of the distribution.

At the same time, the confidence intervals are relatively wide for several quintiles and overlap considerably between the two linguistic groups. This indicates that the differences between national-language speakers and Russian speakers are estimated with limited precision. Taken together, the results suggest that childhood deprivation is associated with a higher likelihood of experiencing multifrailty in later life, while providing limited evidence of systematic differences in this relationship across linguistic groups.

*Figure 4 Correlation between childhood deprivation quantiles and multifrailty by linguistic group in Latvia and Estonia*

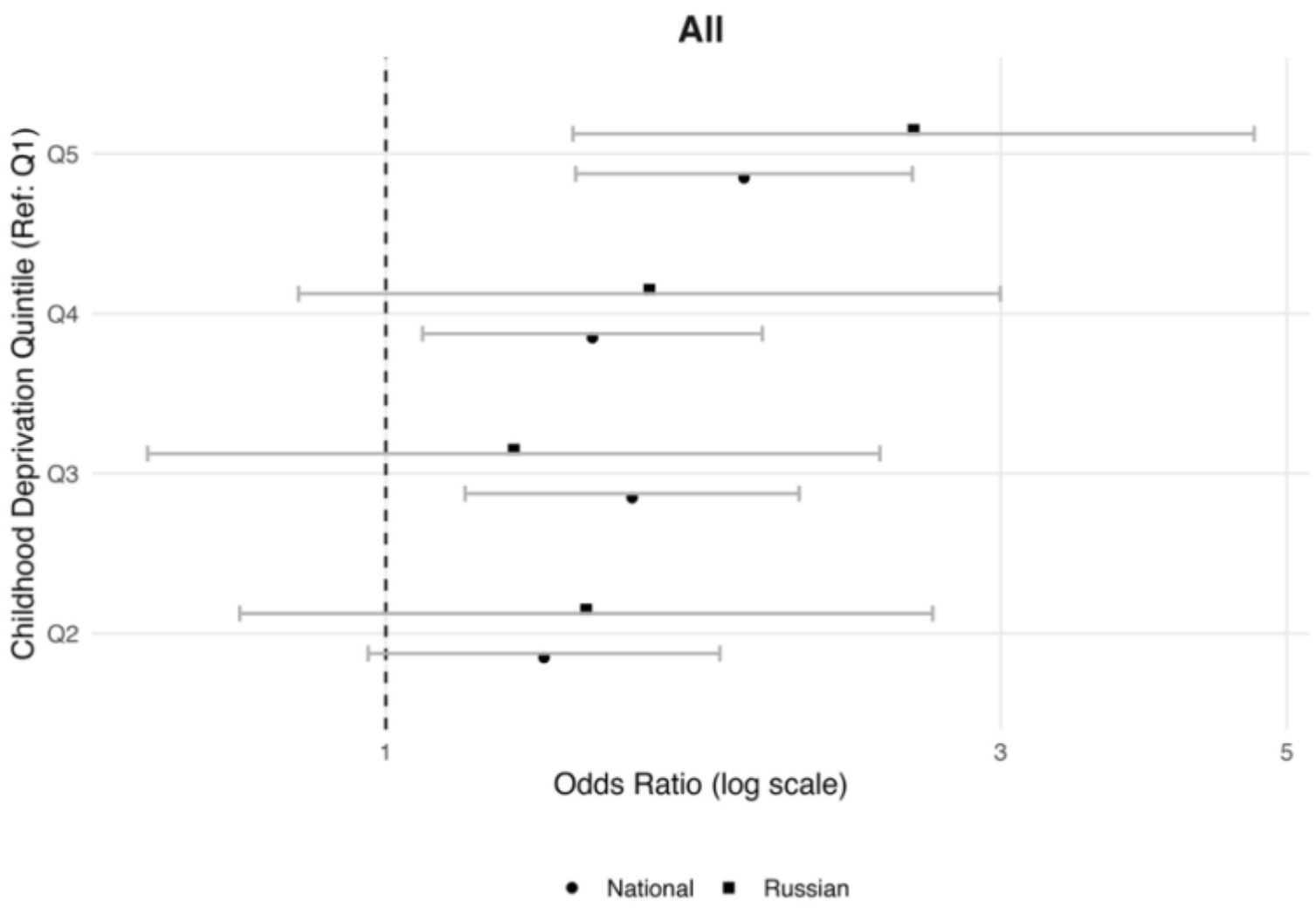

Taken together, the four figures reveal a consistent association between childhood deprivation and multiple measures of health in later life. Across both predicted probabilities and odds ratios, individuals who experienced higher levels of deprivation during childhood generally display elevated risks of poor self-rated health, chronic disease burden, functional limitation, and their combined manifestation as multifrailty. While the magnitude and shape of the gradient vary across outcomes and contexts, the overall pattern indicates that early-life disadvantage remains linked to health vulnerability many decades later.

Overall, the findings suggest a persistent gradient between childhood deprivation and later-life health outcomes. Individuals exposed to higher levels of deprivation during childhood consistently display higher predicted probabilities of poor health across multiple indicators. The similarity of these patterns across countries and linguistic groups indicates that early-life disadvantage remains strongly associated with health outcomes even after individuals experience divergent institutional environments later in life.

Additional robustness checks interacting childhood deprivation with country and linguistic group shows no systematic heteogeneity in the deprivation gradient. The association between childhood depriation and later life health outcomes remains consistent acros countries and linguistic groups[9]

**5. Discussion and conclusion.**

This study examined the relationship between childhood deprivation and health outcomes in later life among older adults in Estonia, Latvia, and Israel, with particular attention to potential linguistic differences in the Baltic states. Using harmonised data from the Survey of Health, Ageing and Retirement in Europe (SHARE) and adopting a life-course perspective, the analysis assessed whether early-life socioeconomic disadvantage is associated with multiple indicators of health in older age. These outcomes include poor self-rated health, chronic disease burden, functional limitations, depression, and multidimensional vulnerability captured through a composite multifrailty indicator. The comparison between Estonia, Latvia, and Israel offers a unique opportunity to examine life-course health persistence among cohorts exposed to common early-life institutions but divergent adult institutional environments.

The findings provide consistent evidence that childhood deprivation is strongly associated with health outcomes in later life. Across multiple empirical specifications, individuals who experienced higher levels of deprivation during childhood face significantly elevated risks of poor self-rated health, chronic disease burden, functional limitations, and depression in older age. The relationship follows a clear gradient, with individuals in higher deprivation quintiles generally displaying the greatest health risks. Importantly, this pattern remains statistically significant even after accounting for adult socioeconomic characteristics. These findings are consistent with a large body of life-course research demonstrating that childhood socioeconomic conditions shape health trajectories through both biological and social mechanisms (Ben-Shlomo and Kuh, 2002; Galobardes et al., 2006; Haas, 2008).

[9] See appendix table A5.

*Figure 5 Conceptual pathways linking childhood deprivation to old-age health outcomes*

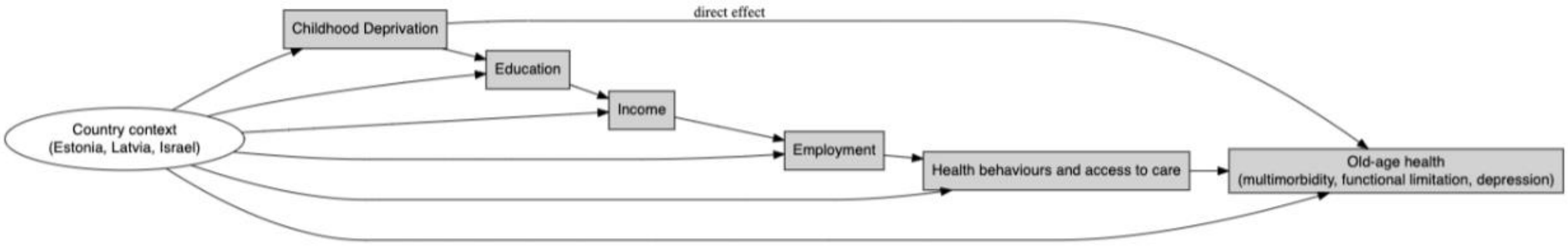

Although some variation across countries is observed, the overall deprivation gradient remains evident in Estonia, Latvia, and Israel. The gradient appears strongest and most monotonic in Estonia, somewhat more moderate in Latvia, and more concentrated at the highest levels of childhood deprivation in Israel. Despite these differences, severe early-life disadvantage is consistently associated with elevated health risks later in life across all three contexts. This pattern suggests that while national institutions and welfare systems may influence overall population health, they appear less able to fully offset inequalities rooted in early-life disadvantage.

Beyond these country-specific differences, the findings also contribute to a broader life-course question concerning the persistence of early disadvantage under conditions of institutional change. The cohorts examined in this study experienced childhood under broadly similar Soviet institutional arrangements but entered adulthood under markedly different socioeconomic systems following the dissolution of the Soviet Union. Estonia and Latvia underwent rapid market transitions and welfare state restructuring, while migrants to Israel were incorporated into a different institutional environment through processes of migration and immigrant integration. This divergence creates a life-course setting in which individuals with comparable early-life exposures subsequently encountered distinct adult opportunity structures. The persistence of the childhood deprivation gradient across these contexts suggests that early-life disadvantage continues to shape health trajectories even when later-life institutional environments differ substantially. In this sense, the findings support the view that inequalities rooted in childhood conditions may endure across institutional transitions rather than being fully offset by subsequent policy environments.

The linguistic dimension adds further nuance to these results. Historical migration patterns and post-independence language policies have produced distinct socioeconomic profiles for Russian-speaking and national-language populations in Estonia and Latvia. Previous research has documented disparities in cognitive impairment (Abuladze et al., 2023), health behaviours and healthcare access (Sepp and Volmer, 2022), and wealth accumulation (Rebane et al., 2024) between these groups. However, the present analysis suggests that the relationship between childhood deprivation and later-life health operates in broadly similar ways across linguistic groups. Severe childhood deprivation is associated with multidimensional vulnerability among both Russian-speaking and national-language respondents. While some differences in magnitude are observed for specific outcomes, there is no systematic evidence that the deprivation–health gradient is structurally stronger for one linguistic group than the other.

At the same time, adult socioeconomic conditions remain an important determinant of health in older age. Financial strain in particular emerges as one of the strongest predictors of adverse health outcomes across the models. Respondents who report difficulty making ends meet face markedly higher risks of functional limitations and multifrailty. These findings support a cumulative inequality perspective in which early-life disadvantage and adult socioeconomic hardship jointly contribute to later-life vulnerability (Dannefer, 2003;

Ferraro et al., 2016). Rather than acting as competing explanations, childhood deprivation and adult financial strain appear to reinforce one another across the life course.

These results also resonate with research on wealth disparities between immigrant and native populations in the Baltic region. For example, Rebane et al. (2024) show that although immigrants and natives began the post-Soviet transition from relatively similar starting points, differences in asset accumulation emerged over time. Such processes likely shape socioeconomic trajectories across adulthood. However, the persistence of childhood deprivation effects after accounting for adult socioeconomic conditions suggests that the roots of later-life health inequality extend further back in the life course. Similar conclusions have been drawn in comparative studies using SHARE data across Europe, which show that childhood health and socioeconomic conditions remain strongly associated with health outcomes in older age (Börsch-Supan et al., 2013; Wahrendorf and Blane, 2015).

Several limitations should be acknowledged. First, childhood conditions are measured retrospectively and may therefore be subject to recall bias. Although retrospective indicators have been widely used in life-course research, respondents' recollections may not perfectly capture early-life circumstances. Second, the analysis is observational and cannot establish causal relationships. Third, the study sample includes only individuals who survived to age 50 or older. Selective mortality may therefore attenuate the estimated association between childhood deprivation and later-life health, as individuals exposed to severe early disadvantage may be less likely to survive into older age. Finally, the study does not distinguish between different migration histories among Russian-speaking populations in Estonia and Latvia. Russian-speaking respondents may include both individuals whose families migrated during the Soviet period and those whose families have lived in the region for several generations, and these groups may differ in their socioeconomic integration and identity formation.

Despite these limitations, the consistency of the deprivation gradient across countries, health outcomes, and linguistic groups strengthens the central conclusion of this study. Early-life socioeconomic disadvantage remains a powerful and persistent predictor of health in later life, even among populations that experienced substantial institutional transformation and migration during adulthood.

In sum, this study demonstrates that childhood deprivation continues to shape health outcomes decades later in Estonia, Latvia, and Israel. Although contemporary disparities are influenced by linguistic stratification and adult socioeconomic conditions, early-life disadvantage remains a key determinant of health in older age. Policies aimed at reducing long-term health inequalities may therefore need to address both present socioeconomic vulnerability and the conditions experienced during childhood. Interventions that improve childhood living conditions—such as investments in nutrition, housing, healthcare, and education—may generate health benefits that extend across the entire life course (Heckman, 2006; Conti and Heckman, 2012). By reducing early-life inequalities, such policies may contribute not only to improved population health but also to more sustainable health systems in ageing societies.

**Acknowledgement**

This paper uses data from SHARELIFE wave 7 (https://doi.org/10.6103/SHARE.w7.900), wave 8 (https://doi.org/10.6103/SHARE.w8.900), and wave 9 (https://doi.org/10.6103/SHARE.w9.900).

**References**

Abuladze, L., Sakkeus, L., Selezneva, E., & Sinyavskaya, O. (2023). Comparing the cognitive functioning of middle-aged and older foreign-origin population in Estonia to host and origin populations. *Frontiers in Public Health*, *11*, 1058578. https://doi.org/10.3389/fpubh.2023.1058578

Alkire, S., & Foster, J. (2011). Counting and multidimensional poverty measurement. *Journal of Public Economics*, *95*(7–8), 476–487. https://doi.org/10.1016/j.jpubeco.2010.11.006

Angelini, V., Klijs, B., Smidt, N., & Mierau, J. O. (2016). Associations between Childhood Parental Mental Health Difficulties and Depressive Symptoms in Late Adulthood: The Influence of Life-Course Socioeconomic, Health and Lifestyle Factors. *PLOS ONE*, *11*(12), e0167703. https://doi.org/10.1371/journal.pone.0167703

Barboza Solís, C., Kelly-Irving, M., Fantin, R., Darnaudéry, M., Torrisani, J., Lang, T., & Delpierre, C. (2015). Adverse childhood experiences and physiological wear-and-tear in midlife: Findings from the 1958 British birth cohort. *Proceedings of the National Academy of Sciences*, *112*(7). https://doi.org/10.1073/pnas.1417325112

Ben-Shlomo, Y. (2002). A life course approach to chronic disease epidemiology: Conceptual models, empirical challenges and interdisciplinary perspectives. *International Journal of Epidemiology*, *31*(2), 285–293. https://doi.org/10.1093/ije/31.2.285

Börsch-Supan, A., Brandt, M., Hunkler, C., Kneip, T., Korbmacher, J., Malter, F., Schaan, B., Stuck, S., & Zuber, S. (2013). Data Resource Profile: The Survey of Health, Ageing and Retirement in Europe (SHARE). *International Journal of Epidemiology*, *42*(4), 992–1001. https://doi.org/10.1093/ije/dyt088

Case, A., Fertig, A., & Paxson, C. (2005). The lasting impact of childhood health and circumstance. *Journal of Health Economics*, *24*(2), 365–389. https://doi.org/10.1016/j.jhealeco.2004.09.008

Case, A., & Paxson, C. (2010). Causes and consequences of early-life health. *Demography*, *47*(Suppl 1), S65–S85. https://doi.org/10.1353/dem.2010.0007

Cheskin, A. (2016). *Russian Speakers in Post-Soviet Latvia: Discursive Identity Strategies*. Edinburgh University Press. https://doi.org/10.1515/9780748697441

Cheval, B., Orsholits, D., Sieber, S., Stringhini, S., Courvoisier, D., Kliegel, M., Boisgontier, M. P., & Cullati, S. (2019). Early-life socioeconomic circumstances explain health differences in old age, but not their evolution over time. *Journal of Epidemiology and Community Health*, *73*(8), 703–711. https://doi.org/10.1136/jech-2019-212110

Conti, G., & Heckman, J. J. (2014). Economics of Child Well-Being. In A. Ben-Arieh, F. Casas, I. Frønes, & J. E. Korbin (Eds), *Handbook of Child Well-Being* (pp. 363–401). Springer Netherlands. https://doi.org/10.1007/978-90-481-9063-8_21

Cook, L. J. (2013). *Postcommunist Welfare States: Reform Politics in Russia and Eastern Europe*. Cornell University Press. https://doi.org/10.7591/9780801460098

Dannefer, D. (2003). Cumulative Advantage/Disadvantage and the Life Course: Cross-Fertilizing Age and Social Science Theory. *The Journals of Gerontology Series B: Psychological Sciences and Social Sciences*, *58*(6), S327–S337. https://doi.org/10.1093/geronb/58.6.S327

Eikemo, T. A., Bambra, C., Joyce, K., & Dahl, E. (2008). Welfare state regimes and income-related health inequalities: A comparison of 23 European countries. *European Journal of Public Health*, *18*(6), 593–599. https://doi.org/10.1093/eurpub/ckn092

Ferraro, K. F., Schafer, M. H., & Wilkinson, L. R. (2016). Childhood Disadvantage and Health Problems in Middle and Later Life: Early Imprints on Physical Health? *American Sociological Review*, *81*(1), 107–133. https://doi.org/10.1177/0003122415619617

Galobardes, B. (2004). Childhood Socioeconomic Circumstances and Cause-specific Mortality in Adulthood: Systematic Review and Interpretation. *Epidemiologic Reviews*, *26*(1), 7–21. https://doi.org/10.1093/epirev/mxh008

Gorodzeisky, A., & Semyonov, M. (2011). Two Dimensions to Economic Incorporation: Soviet Immigrants in the Israeli Labour Market. *Journal of Ethnic and Migration Studies*, *37*(7), 1059–1077. https://doi.org/10.1080/1369183X.2011.572483

Haas, S. (2008). Trajectories of functional health: The 'long arm' of childhood health and socioeconomic factors. *Social Science & Medicine*, *66*(4), 849–861. https://doi.org/10.1016/j.socscimed.2007.11.004

Havari, E., & Mazzonna, F. (2015). Can We Trust Older People's Statements on Their Childhood Circumstances? Evidence from SHARELIFE. *European Journal of Population*, *31*(3), 233–257. https://doi.org/10.1007/s10680-014-9332-y

Heckman, J. J. (2006). Skill Formation and the Economics of Investing in Disadvantaged Children. *Science*, *312*(5782), 1900–1902. https://doi.org/10.1126/science.1128898

Hughes, K., Bellis, M. A., Hardcastle, K. A., Sethi, D., Butchart, A., Mikton, C., Jones, L., & Dunne, M. P. (2017). The effect of multiple adverse childhood experiences on health: A systematic review and meta-analysis. *The Lancet Public Health*, *2*(8), e356–e366. https://doi.org/10.1016/S2468-2667(17)30118-4

Lacey, R. E., Bartley, M., Kelly-Irving, M., Bevilacqua, L., Iob, E., Kelly, Y., & Howe, L. D. (2020). Adverse childhood experiences and early life inflammation in the Avon longitudinal study of parents and children. *Psychoneuroendocrinology*, *122*, 104914. https://doi.org/10.1016/j.psyneuen.2020.104914

Leshem, E. (2008). Being an Israeli: Immigrants from the Former Soviet Union in Israel, fifteen years later. *Journal of Israeli History*, *27*(1), 29–49. https://doi.org/10.1080/13531040801902716

Rebane, L., Kukk, M., & Rõõm, T. (2024). *Wealth disparities between elderly immigrants and natives: A study of Estonia and Latvia*. https://doi.org/10.23656/25045520/042024/0214

Remennick, L. (2014). *Russian Israelis: Social Mobility, Politics and Culture*. Taylor and Francis.

Scherpenzeel, A., Axt, K., Bergmann, M., Douhou, S., Oepen, A., Sand, G., Schuller, K., Stuck, S., Wagner, M., & Börsch-Supan, A. (2020). Collecting survey data among the 50+ population during the COVID-19 outbreak: The Survey of Health, Ageing and Retirement in Europe (SHARE). *Survey Research Methods*, 217-221 Pages. https://doi.org/10.18148/SRM/2020.V14I2.7738

Sepp, K., & Volmer, D. (2022). Experiences and Expectations of Ethnic Minorities and Majorities towards Community Pharmacy Medicines-Related Services in Estonia. *International Journal of Environmental Research and Public Health*, *19*(8), 4755. https://doi.org/10.3390/ijerph19084755

SHARE-ERIC. (2024a). *Survey of Health, Ageing and Retirement in Europe (SHARE) Wave 7* (Version 9.0.0) [Data set]. SHARE-ERIC. https://doi.org/10.6103/SHARE.W7.900

SHARE-ERIC. (2024b). *Survey of Health, Ageing and Retirement in Europe (SHARE) Wave 8* (Version 9.0.0) [Data set]. SHARE-ERIC. https://doi.org/10.6103/SHARE.W8.900

SHARE-ERIC. (2024c). *Survey of Health, Ageing and Retirement in Europe (SHARE) Wave 9* (Version 9.0.0) [Data set]. SHARE-ERIC. https://doi.org/10.6103/SHARE.W9.900

Simonyan, R. H. (2022). The Russian-speaking Diaspora in the Baltic States: A socio-cultural aspect. *Baltic Region*, *14*(2), 144–157. https://doi.org/10.5922/2079-8555-2022-2-9

Wahrendorf, M., & Blane, D. (2015). Does labour market disadvantage help to explain why childhood circumstances are related to quality of life at older ages? Results from SHARE. *Aging & Mental Health*, *19*(7), 584–594. https://doi.org/10.1080/13607863.2014.938604

Weber, G. (2018). Share: A Data Set for Ageing Research. *Journal of Public Health Research*, *7*(1), jphr.2018.1397. https://doi.org/10.4081/jphr.2018.1397

Wu, S., Wang, R., Zhao, Y., Ma, X., Wu, M., Yan, X., & He, J. (2013). The relationship between self-rated health and objective health status: A population-based study. *BMC Public Health*, *13*(1), 320. https://doi.org/10.1186/1471-2458-13-320

Appendix

*Table A 1 Association between health outcomes and childhood deprivation, with additional controls (OR)*

| VARIABLES | (a) Poor self-rated health | Chronic disease burden | Functional limitation | Depression | (b) Poor self-rated health AND Chronic diseases burden | Poor self-rated health AND Chronic diseases burden | Chronic diseases burden AND Functional limitation | All |
|---|---|---|---|---|---|---|---|---|
| Total childhood deprivation score (Ref. Quintile 1) | | | | | | | | |
| Quintile 2 | 1.207** | 1.238** | 1.328** | 1.186 | 1.348** | 1.294** | 1.188* | 1.319** |
| | (0.104) | (0.107) | (0.157) | (0.136) | (0.169) | (0.170) | (0.108) | (0.178) |
| Quintile 3 | 1.477*** | 1.277*** | 1.446*** | 1.131 | 1.494*** | 1.437*** | 1.325*** | 1.486*** |
| | (0.130) | (0.109) | (0.165) | (0.131) | (0.179) | (0.180) | (0.119) | (0.191) |
| Quintile 4 | 1.550*** | 1.374*** | 1.527*** | 1.268** | 1.557*** | 1.553*** | 1.392*** | 1.569*** |
| | (0.147) | (0.122) | (0.175) | (0.148) | (0.188) | (0.195) | (0.130) | (0.203) |
| Quintile 5 | 1.826*** | 1.650*** | 2.101*** | 1.823*** | 2.128*** | 2.150*** | 1.700*** | 2.164*** |
| | (0.185) | (0.153) | (0.241) | (0.217) | (0.255) | (0.265) | (0.161) | (0.274) |
| Demographic indicators | | | | | | | | |
| Is a female | 0.954 | 1.194*** | 1.496*** | 1.744*** | 1.469*** | 1.578*** | 1.218*** | 1.573*** |
| | (0.060) | (0.071) | (0.111) | (0.138) | (0.113) | (0.124) | (0.075) | (0.127) |
| Age | 1.018 | 1.061*** | 0.951*** | 0.927*** | 0.952*** | 0.979 | 1.047*** | 0.974 |
| | (0.014) | (0.014) | (0.016) | (0.019) | (0.016) | (0.017) | (0.014) | (0.018) |
| Age squared | 1.082** | 0.968 | 1.298*** | 1.215*** | 1.280*** | 1.212*** | 0.999 | 1.214*** |
| | (0.034) | (0.027) | (0.043) | (0.049) | (0.044) | (0.043) | (0.028) | (0.044) |
| Education level (*Ref. None*) | | | | | | | | |
| Primary education | 0.391 | 1.374 | 0.842 | 0.976 | 0.802 | 1.050 | 0.816 | 1.048 |
| | (0.257) | (0.583) | (0.330) | (0.635) | (0.319) | (0.411) | (0.354) | (0.405) |
| Secondary education | 0.246** | 1.117 | 0.650 | 0.833 | 0.591 | 0.809 | 0.610 | 0.784 |
| | (0.161) | (0.471) | (0.253) | (0.541) | (0.234) | (0.315) | (0.264) | (0.302) |
| Tertiary education | 0.184*** | 1.012 | 0.469* | 0.814 | 0.424** | 0.595 | 0.510 | 0.572 |
| | (0.121) | (0.428) | (0.184) | (0.530) | (0.169) | (0.234) | (0.221) | (0.222) |
| **Country** (*Ref. Estonia*) | | | | | | | | |
| Estonia | 0.764*** | 0.943 | 0.696*** | 0.629*** | 0.707*** | 0.721*** | 0.930 | 0.729*** |
| | (0.056) | (0.065) | (0.059) | (0.059) | (0.061) | (0.065) | (0.066) | (0.066) |
| Latvia | 0.171*** | 0.733*** | 0.697*** | 1.004 | 0.585*** | 0.646*** | 0.338*** | 0.571*** |
| | (0.015) | (0.061) | (0.074) | (0.120) | (0.065) | (0.072) | (0.030) | (0.067) |
| Employment status (*ref. Retired)* | | | | | | | | |
| Employed or self-employed | 0.618*** | 0.686*** | 0.521*** | 0.688*** | 0.426*** | 0.453*** | 0.610*** | 0.401*** |
| | (0.049) | (0.052) | (0.057) | (0.078) | (0.050) | (0.055) | (0.050) | (0.052) |
| Others | 2.030*** | 1.561*** | 2.336*** | 1.212 | 2.265*** | 2.502*** | 1.799*** | 2.362*** |
| | (0.220) | (0.150) | (0.254) | (0.171) | (0.252) | (0.290) | (0.177) | (0.278) |
| Household indicators | | | | | | | | |
| Is living with a partner | 0.856* | 0.956 | 0.857* | 0.870 | 0.879 | 0.859 | 0.957 | 0.879 |
| | (0.069) | (0.069) | (0.075) | (0.081) | (0.080) | (0.080) | (0.071) | (0.083) |
| Lives in Urban area | 1.013 | 1.155** | 1.004 | 1.067 | 1.013 | 1.088 | 1.133* | 1.074 |
| | (0.073) | (0.078) | (0.079) | (0.090) | (0.082) | (0.092) | (0.079) | (0.093) |
| Household size | 0.863*** | 0.873*** | 0.844*** | 0.739*** | 0.834*** | 0.807*** | 0.876** | 0.791*** |
| | (0.046) | (0.045) | (0.053) | (0.063) | (0.055) | (0.058) | (0.048) | (0.059) |
| Household income quartile (*ref. First quartile*) | | | | | | | | |
| Second quartile | 0.860* | 0.847** | 0.868 | 0.871 | 0.875 | 0.864 | 0.835** | 0.872 |
| | (0.078) | (0.069) | (0.083) | (0.116) | (0.086) | (0.089) | (0.069) | (0.091) |
| Third quartile | 0.718*** | 0.791** | 0.774** | 0.634*** | 0.763** | 0.760** | 0.730*** | 0.755** |
| | (0.072) | (0.074) | (0.091) | (0.093) | (0.094) | (0.099) | (0.072) | (0.102) |
| Fourth quartile | 0.542*** | 0.714*** | 0.630*** | 0.489*** | 0.639*** | 0.605*** | 0.666*** | 0.600*** |
| | (0.075) | (0.087) | (0.090) | (0.093) | (0.094) | (0.094) | (0.084) | (0.096) |
| Difficulty in making ends meet: Yes | 1.863*** | 1.526*** | 2.226*** | 1.699*** | 2.244*** | 2.149*** | 1.674*** | 2.143*** |
| | (0.102) | (0.078) | (0.140) | (0.120) | (0.146) | (0.144) | (0.087) | (0.146) |
| The respondent was born abroad | 1.413*** | 1.404*** | 1.045 | 1.315*** | 1.140* | 1.245*** | 1.478*** | 1.315*** |
| | (0.108) | (0.096) | (0.081) | (0.114) | (0.090) | (0.101) | (0.102) | (0.108) |
| Wave (ref. 7) | | | | | | | | |
| Wave 8 | 1.011 | 1.071 | 1.137** | | 1.168*** | 1.166** | 1.078 | 1.218*** |
| | (0.053) | (0.050) | (0.064) | | (0.068) | (0.072) | (0.052) | (0.076) |
| Wave 9 | 0.963 | 1.253*** | 0.774*** | 1.122** | 0.788*** | 0.905 | 1.136** | 0.916 |
| | (0.052) | (0.061) | (0.049) | (0.062) | (0.051) | (0.061) | (0.057) | (0.063) |
| Constant | 2.459 | 0.014*** | 2.969 | 50.164*** | 2.930 | 0.296 | 0.044*** | 0.411 |
| | (2.632) | (0.013) | (3.288) | (72.431) | (3.362) | (0.354) | (0.042) | (0.499) |
| Observations | 10,079 | 10,079 | 10,079 | 5,318 | 10,079 | 10,079 | 10,079 | 10,079 |
| Log-likelihood | -5356 | -6255 | -4420 | -3081 | -4223 | -3946 | -5967 | -3837 |
| Pseudo R2 | 0.188 | 0.103 | 0.192 | 0.0773 | 0.197 | 0.193 | 0.135 | 0.194 |

Robust standard errors in parentheses (*** p<0.01, ** p<0.05, * p<0.1)

*Table A 2 Association between health outcome and childhood deprivation, with additional controls, by countries (OR)*

| | Estonia | | | | Latvia | | | | Israel | | | |
|---|---|---|---|---|---|---|---|---|---|---|---|---|
| | Poor self-rated health | Chronic disease burden | Functional limitation | Depression | Poor self-rated health | Chronic disease burden | Functional limitation | Depression | Poor self-rated health | Chronic disease burden | Functional limitation | Depression |
| Total childhood deprivation score (Ref. Quintile 1) | | | | | | | | | | | | |
| Quintile 2 | 1.351*** | 1.359*** | 1.452** | 1.110 | 1.075 | 0.928 | 1.018 | 1.381 | 0.852 | 1.329 | 0.869 | 1.533 |
| | (0.150) | (0.148) | (0.210) | (0.154) | (0.186) | (0.170) | (0.262) | (0.351) | (0.219) | (0.294) | (0.291) | (0.532) |
| Quintile 3 | 1.443*** | 1.355*** | 1.635*** | 1.072 | 1.700*** | 1.297 | 0.949 | 1.357 | 1.491* | 1.064 | 1.154 | 1.226 |
| | (0.162) | (0.146) | (0.227) | (0.148) | (0.331) | (0.250) | (0.242) | (0.361) | (0.329) | (0.228) | (0.380) | (0.438) |
| Quintile 4 | 1.501*** | 1.413*** | 1.609*** | 1.194 | 1.403* | 1.356 | 0.809 | 1.251 | 1.974*** | 1.232 | 2.157** | 1.940* |
| | (0.183) | (0.159) | (0.227) | (0.168) | (0.285) | (0.266) | (0.204) | (0.339) | (0.451) | (0.261) | (0.648) | (0.683) |
| Quintile 5 | 1.700*** | 1.876*** | 2.034*** | 1.854*** | 1.533** | 1.322 | 1.263 | 1.621* | 2.932*** | 1.330 | 4.063*** | 1.537 |
| | (0.225) | (0.225) | (0.290) | (0.270) | (0.314) | (0.256) | (0.316) | (0.446) | (0.667) | (0.306) | (1.178) | (0.502) |
| Demographic indicators | | | | | | | | | | | | |
| Is a female | 0.811** | 1.200** | 1.561*** | 1.707*** | 1.446*** | 1.609*** | 1.847*** | 1.875*** | 1.066 | 0.832 | 1.063 | 2.371*** |
| | (0.069) | (0.092) | (0.142) | (0.164) | (0.186) | (0.201) | (0.330) | (0.345) | (0.172) | (0.126) | (0.220) | (0.601) |
| Age | 0.999 | 1.044** | 0.935*** | 0.914*** | 1.077*** | 1.042 | 0.969 | 0.998 | 1.023 | 1.129*** | 0.881*** | 0.907 |
| | (0.019) | (0.017) | (0.019) | (0.023) | (0.030) | (0.029) | (0.042) | (0.056) | (0.039) | (0.043) | (0.040) | (0.068) |
| Age squared | 1.133*** | 0.992 | 1.334*** | 1.247*** | 0.971 | 1.016 | 1.278*** | 0.987 | 1.052 | 0.863* | 1.513*** | 1.359** |
| | (0.048) | (0.034) | (0.054) | (0.061) | (0.066) | (0.065) | (0.113) | (0.112) | (0.085) | (0.070) | (0.147) | (0.205) |
| Education level (*Ref. None*) | | | | | | | | | | | | |
| Primary education | 1.753 | 1.278 | 0.887 | 0.881 | 1.734* | 2.746 | 0.407 | 1.553 | 0.358 | 1.041 | 1.573 | 1.587 |
| | (1.069) | (0.828) | (0.634) | (0.847) | (0.494) | (2.128) | (0.397) | (0.474) | (0.279) | (0.738) | (0.945) | (1.354) |
| Secondary education | 1.105 | 1.067 | 0.719 | 0.779 | 1.272* | 1.879 | 0.301 | 1.201 | 0.202** | 0.781 | 0.808 | 0.717 |
| | (0.666) | (0.688) | (0.513) | (0.747) | (0.172) | (1.439) | (0.290) | (0.222) | (0.156) | (0.551) | (0.490) | (0.607) |
| Tertiary education | 0.766 | 0.862 | 0.458 | 0.724 | - | 1.914 | 0.197* | - | 0.160** | 0.939 | 0.987 | 1.449 |
| | (0.463) | (0.557) | (0.328) | (0.695) | | (1.476) | (0.192) | | (0.123) | (0.662) | (0.596) | (1.235) |
| Employment status (*ref. Retired)* | | | | | | | | | | | | |
| Employed or self-employed | 0.678*** | 0.707*** | 0.545*** | 0.678*** | 0.626*** | 0.547*** | 0.385*** | 0.589 | 0.415*** | 0.610*** | 0.347*** | 0.862 |
| | (0.070) | (0.071) | (0.071) | (0.092) | (0.110) | (0.098) | (0.119) | (0.199) | (0.086) | (0.104) | (0.115) | (0.264) |
| Others | 2.207*** | 1.491*** | 2.066*** | 0.900 | 2.858*** | 1.309 | 2.377*** | 1.479 | 1.447** | 1.991*** | 2.931*** | 1.566* |
| | (0.386) | (0.213) | (0.318) | (0.180) | (0.710) | (0.289) | (0.695) | (0.557) | (0.272) | (0.370) | (0.651) | (0.420) |
| Household indicators | | | | | | | | | | | | |
| Is living with a partner | 0.836* | 0.918 | 0.888 | 0.833* | 1.242 | 1.506*** | 0.907 | 1.085 | 0.606** | 0.700 | 0.743 | 0.859 |
| | (0.084) | (0.081) | (0.094) | (0.092) | (0.210) | (0.233) | (0.172) | (0.260) | (0.146) | (0.155) | (0.211) | (0.251) |
| Lives in Urban area | 1.003 | 1.121 | 0.950 | 1.124 | 1.069 | 1.295** | 1.363* | 0.892 | 1.004 | 1.452 | 0.745 | 2.729 |
| | (0.090) | (0.093) | (0.086) | (0.110) | (0.139) | (0.159) | (0.220) | (0.155) | (0.517) | (0.647) | (0.449) | (2.821) |
| Household size | 0.905 | 0.910 | 0.882 | 0.758** | 0.785** | 0.887 | 0.842 | 0.711* | 0.788* | 0.701** | 0.656** | 0.587* |
| | (0.065) | (0.060) | (0.068) | (0.082) | (0.080) | (0.091) | (0.108) | (0.127) | (0.102) | (0.104) | (0.110) | (0.173) |
| Household income quartile (*ref. First quartile*) | | | | | | | | | | | | |
| Second quartile | 1.022 | 0.910 | 0.871 | 0.834 | 0.755 | 0.795 | 1.025 | 0.947 | 0.574** | 0.681 | 0.695 | 0.602 |
| | (0.118) | (0.091) | (0.100) | (0.134) | (0.146) | (0.143) | (0.222) | (0.282) | (0.144) | (0.164) | (0.214) | (0.296) |
| Third quartile | 0.794* | 0.858 | 0.768* | 0.586*** | 0.687* | 0.764 | 1.096 | 0.874 | 0.521** | 0.582** | 0.674 | 0.459 |
| | (0.101) | (0.101) | (0.113) | (0.106) | (0.150) | (0.154) | (0.279) | (0.275) | (0.136) | (0.144) | (0.210) | (0.226) |
| Fourth quartile | 0.662** | 0.755* | 0.650** | 0.459*** | 0.539** | 0.983 | 0.826 | 0.820 | 0.266*** | 0.359*** | 0.420** | 0.172*** |
| | (0.116) | (0.115) | (0.115) | (0.109) | (0.160) | (0.248) | (0.254) | (0.333) | (0.099) | (0.128) | (0.174) | (0.115) |
| Difficulty in making ends meet: Yes | 1.022 | 0.910 | 0.871 | 0.834 | 1.845*** | 1.547*** | 1.656*** | 1.797*** | 1.656*** | 1.243 | 2.004*** | 1.754** |
| | (0.118) | (0.091) | (0.100) | (0.134) | (0.218) | (0.173) | (0.249) | (0.314) | (0.255) | (0.186) | (0.386) | (0.412) |
| The respondent | 0.794* | 0.858 | 0.768* | 0.586*** | 0.834 | 0.976 | 0.736* | 1.148 | 1.637*** | 1.564*** | 1.373 | 1.623** |
| was born abroad | (0.101) | (0.101) | (0.113) | (0.106) | (0.142) | (0.147) | (0.134) | (0.240) | (0.262) | (0.231) | (0.273) | (0.377) |
| | | | | | | | | | | | | |
| Wave (ref. 7) | | | | | | | | | | | | |
| Wave 8 | 0.843*** | 0.978 | 1.126* | | 1.286** | 1.089 | 0.890 | | 1.467*** | 1.528*** | 1.688*** | |
| | (0.055) | (0.055) | (0.076) | | (0.155) | (0.120) | (0.126) | | (0.199) | (0.202) | (0.293) | |
| Wave 9 | 0.846** | 1.160** | 0.726*** | 1.282*** | 1.141 | 1.237* | 0.867 | 0.792* | 1.202 | 2.167*** | 0.873 | 0.775 |
| | (0.057) | (0.068) | (0.055) | (0.083) | (0.132) | (0.136) | (0.121) | (0.111) | (0.222) | (0.360) | (0.214) | (0.147) |
| | | | | | | | | | | | | |
| Constant | 1.618 | 0.039*** | 6.998 | 129.634*** | 0.010*** | 0.016** | 1.650 | 0.308 | 0.725 | 0.001*** | 296.892* | 57.981 |
| | (2.092) | (0.047) | (9.940) | (234.903) | (0.016) | (0.029) | (4.765) | (1.072) | (1.890) | (0.002) | (924.699) | (279.868) |
| | | | | | | | | | | | | |
| Observations | 6,487 | 6,487 | 6,487 | 3,683 | 2,100 | 2,107 | 2,107 | 1,046 | 1,485 | 1,485 | 1,485 | 589 |
| Log-likelihood | -3389 | -4048 | -3012 | -2178 | -1143 | -1280 | -858.4 | -556 | -759.1 | -870.1 | -494 | -316 |
| Pseudo R2 | 0.143 | 0.0957 | 0.185 | 0.0754 | 0.165 | 0.123 | 0.181 | 0.0725 | 0.216 | 0.153 | 0.292 | 0.150 |

Robust standard errors in parentheses (*** p<0.01, ** p<0.05, * p<0.1)

*Table A 3 Association between multifrailty and childhood deprivation, with additional controls, by countries (OR)*

| | Estonia | | | | Latvia | | | | Israel | | | |
|---|---|---|---|---|---|---|---|---|---|---|---|---|
| | Poor self-rated health AND Chronic diseases burden | Poor self-rated health AND Chronic diseases burden | Chronic diseases burden AND Functional limitation | All | Poor self-rated health AND Chronic diseases burden | Poor self-rated health AND Chronic diseases burden | Chronic diseases burden AND Functional limitation | All | Poor self-rated health AND Chronic diseases burden | Poor self-rated health AND Chronic diseases burden | Chronic diseases burden AND Functional limitation | All |
| Total childhood deprivation score (Ref. Quintile 1) | | | | | | | | | | | | |
| Quintile 2 | 1.446** | 1.473** | 1.294** | 1.446** | 0.992 | 0.907 | 1.030 | 0.876 | 1.043 | 0.794 | 0.870 | 1.091 |
| | (0.220) | (0.236) | (0.147) | (0.235) | (0.259) | (0.265) | (0.192) | (0.257) | (0.415) | (0.294) | (0.243) | (0.463) |
| Quintile 3 | 1.615*** | 1.665*** | 1.303** | 1.620*** | 0.937 | 0.948 | 1.436* | 0.925 | 1.906* | 0.911 | 1.349 | 1.656 |
| | (0.236) | (0.256) | (0.145) | (0.253) | (0.244) | (0.261) | (0.278) | (0.254) | (0.707) | (0.320) | (0.333) | (0.652) |
| Quintile 4 | 1.607*** | 1.610*** | 1.301** | 1.545*** | 0.764 | 0.948 | 1.399* | 0.906 | 3.152*** | 1.739* | 1.760** | 2.815*** |
| | (0.238) | (0.250) | (0.151) | (0.244) | (0.195) | (0.257) | (0.281) | (0.246) | (1.095) | (0.574) | (0.432) | (1.055) |
| Quintile 5 | 2.042*** | 2.245*** | 1.777*** | 2.146*** | 1.147 | 1.382 | 1.258 | 1.229 | 5.910*** | 2.435*** | 2.096*** | 4.441*** |
| | (0.304) | (0.345) | (0.215) | (0.335) | (0.291) | (0.372) | (0.240) | (0.333) | (1.922) | (0.767) | (0.514) | (1.573) |
| Demographic indicators | | | | | | | | | | | | |
| Is a female | 1.488*** | 1.550*** | 1.140* | 1.535*** | 1.897*** | 2.332*** | 1.693*** | 2.265*** | 1.182 | 1.263 | 1.068 | 1.347 |
| | (0.140) | (0.149) | (0.089) | (0.151) | (0.339) | (0.439) | (0.212) | (0.426) | (0.268) | (0.285) | (0.188) | (0.331) |
| Age | 0.933*** | 0.950** | 1.024 | 0.945** | 0.994 | 1.012 | 1.092*** | 1.018 | 0.887** | 0.946 | 1.082* | 0.948 |
| | (0.020) | (0.021) | (0.018) | (0.021) | (0.043) | (0.046) | (0.031) | (0.046) | (0.045) | (0.049) | (0.048) | (0.053) |
| Age squared | 1.325*** | 1.269*** | 1.038 | 1.274*** | 1.209** | 1.174* | 0.932 | 1.164 | 1.451*** | 1.308*** | 0.928 | 1.257** |
| | (0.056) | (0.055) | (0.037) | (0.056) | (0.108) | (0.108) | (0.058) | (0.108) | (0.151) | (0.136) | (0.082) | (0.138) |
| Education level (*Ref. None*) | | | | | | | | | | | | |
| Primary education | 0.800 | 1.504 | 1.081 | 1.428 | 0.368 | 0.529 | 1.951 | 0.489 | 1.526 | 1.714 | 0.686 | 1.718 |
| | (0.564) | (0.703) | (0.704) | (0.670) | (0.353) | (0.465) | (1.519) | (0.419) | (0.891) | (1.084) | (0.479) | (1.010) |
| Secondary education | 0.620 | 1.235 | 0.855 | 1.133 | 0.253 | 0.408 | 1.422 | 0.371 | 0.740 | 0.725 | 0.376 | 0.723 |
| | (0.436) | (0.576) | (0.554) | (0.531) | (0.241) | (0.354) | (1.096) | (0.314) | (0.435) | (0.471) | (0.260) | (0.437) |
| Tertiary education | 0.400 | 0.810 | 0.647 | 0.754 | 0.168* | 0.297 | 1.400 | 0.265 | 0.821 | 0.806 | 0.355 | 0.722 |
| | (0.283) | (0.381) | (0.420) | (0.356) | (0.161) | (0.261) | (1.086) | (0.228) | (0.482) | (0.519) | (0.245) | (0.432) |
| Employment status (*ref. Retired)* | | | | | | | | | | | | |
| Employed or self-employed | (0.118) | (0.116) | (0.114) | (0.118) | 0.390*** | 0.476** | 0.633** | 0.489** | 0.238*** | 0.306*** | 0.385*** | 0.213*** |
| | 0.442*** | 0.430*** | 0.615*** | 0.379*** | (0.130) | (0.159) | (0.118) | (0.168) | (0.095) | (0.122) | (0.095) | (0.105) |
| Others | (0.061) | (0.062) | (0.064) | (0.057) | 2.814*** | 2.720*** | 2.029*** | 2.952*** | 2.484*** | 3.236*** | 1.714*** | 2.691*** |
| | 2.032*** | 2.107*** | 1.721*** | 2.024*** | (0.817) | (0.864) | (0.464) | (0.943) | (0.572) | (0.754) | (0.329) | (0.653) |
| Household indicators | | | | | | | | | | | | |
| Is living with a partner | 0.869 | 0.818* | 0.901 | 0.818* | 1.074 | 1.131 | 1.441** | 1.182 | 0.728 | 0.921 | 0.792 | 1.032 |
| | (0.096) | (0.091) | (0.080) | (0.093) | (0.209) | (0.231) | (0.230) | (0.244) | (0.224) | (0.288) | (0.193) | (0.340) |
| Lives in Urban area | 0.959 | 0.991 | 1.097 | 0.983 | 1.400** | 1.506** | 1.253* | 1.477** | 0.756 | 3.302 | 1.374 | 3.095 |
| | (0.090) | (0.097) | (0.092) | (0.098) | (0.229) | (0.267) | (0.156) | (0.263) | (0.485) | (2.486) | (0.930) | (2.384) |
| Household size | 0.864* | 0.838** | 0.907 | 0.811** | 0.812 | 0.863 | 0.883 | 0.849 | 0.660** | 0.507*** | 0.606*** | 0.497*** |
| | (0.071) | (0.073) | (0.063) | (0.074) | (0.114) | (0.134) | (0.097) | (0.138) | (0.112) | (0.103) | (0.094) | (0.108) |
| Household income quartile (*ref. First quartile*) | | | | | | | | | | | | |
| Second quartile | 0.903 | 0.908 | 0.913 | 0.921 | 0.973 | 0.995 | 0.802 | 0.980 | 0.648 | 0.502** | 0.495*** | 0.502* |
| | (0.106) | (0.111) | (0.092) | (0.115) | (0.215) | (0.239) | (0.147) | (0.240) | (0.217) | (0.167) | (0.129) | (0.177) |
| Third quartile | 0.724** | 0.747* | 0.793* | 0.718** | 1.109 | 1.117 | 0.748 | 1.166 | 0.771 | 0.538* | 0.418*** | 0.531* |
| | (0.112) | (0.122) | (0.097) | (0.121) | (0.286) | (0.318) | (0.158) | (0.334) | (0.262) | (0.181) | (0.115) | (0.190) |
| Fourth quartile | 0.645** | 0.605*** | 0.731** | 0.597*** | 0.867 | 0.988 | 0.869 | 0.984 | 0.403** | 0.268*** | 0.219*** | 0.261*** |
| | (0.118) | (0.116) | (0.114) | (0.118) | (0.271) | (0.342) | (0.230) | (0.342) | (0.181) | (0.126) | (0.086) | (0.132) |
| Difficulty in making ends meet: Yes | 2.406*** | 2.236*** | 1.659*** | 2.238*** | 1.744*** | 1.888*** | 1.664*** | 1.879*** | 1.895*** | 1.714*** | 1.622*** | 1.638** |
| | (0.184) | (0.176) | (0.106) | (0.179) | (0.269) | (0.318) | (0.191) | (0.320) | (0.382) | (0.353) | (0.262) | (0.353) |
| The respondent was born abroad | 1.091 | 1.253** | 1.574*** | 1.307** | 0.759 | 0.870 | 0.862 | 0.895 | 1.941*** | 1.655** | 1.918*** | 2.069*** |
| | (0.111) | (0.129) | (0.143) | (0.136) | (0.141) | (0.167) | (0.127) | (0.173) | (0.405) | (0.346) | (0.321) | (0.459) |
| Wave (ref. 7) | | | | | | | | | | | | |
| Wave 8 | 1.148** | 1.124 | 0.955 | 1.146* | 0.941 | 0.970 | 1.151 | 1.041 | 1.742*** | 1.877*** | 1.779*** | 2.228*** |
| | (0.079) | (0.081) | (0.054) | (0.084) | (0.137) | (0.153) | (0.131) | (0.167) | (0.325) | (0.369) | (0.265) | (0.453) |
| Wave 9 | 0.729*** | 0.827** | 1.037 | 0.822** | 0.896 | 1.053 | 1.230* | 1.076 | 0.981 | 1.270 | 1.570** | 1.405 |
| | (0.056) | (0.066) | (0.062) | (0.067) | (0.128) | (0.155) | (0.136) | (0.161) | (0.244) | (0.317) | (0.299) | (0.352) |
| Constant | 9.565 | 1.369 | 0.131 | 2.226 | 0.350 | 0.029 | 0.001*** | 0.022 | 114.539 | 1.242 | 0.008 | 0.675 |
| | (13.970) | (1.944) | (0.163) | (3.192) | (1.031) | (0.087) | (0.002) | (0.067) | (397.148) | (4.382) | (0.025) | (2.563) |
| Observations | 6,487 | 6,487 | 6,487 | 6,487 | 2,107 | 2,107 | 2,107 | 2,107 | 1,485 | 1,485 | 1,485 | 1,485 |
| Log-likelihood | -2894 | -2716 | -3980 | -2654 | -830.3 | -749.8 | -1248 | -737 | -441.8 | -432.8 | -677.9 | -396.4 |
| Pseudo R2 | 0.190 | 0.184 | 0.113 | 0.186 | 0.181 | 0.188 | 0.136 | 0.185 | 0.300 | 0.293 | 0.214 | 0.300 |

Robust standard errors in parentheses (*** p<0.01, ** p<0.05, * p<0.1)

*Table A 4 Association between health outcome and childhood deprivation, by language spoken (Only Estonia and Latvia) (OR)*

| | National language speaker (Estonian/Latvian) | | | | | | | | Russian speaker | | | | | | | |
|---|---|---|---|---|---|---|---|---|---|---|---|---|---|---|---|---|
| VARIABLES | Subjective health: Poor | Numbers of chronic disease | Numbers of IADL restriction | EURO depression | Subjective health: Poor AND any ADL/IADL restrictions | >1 of chronic diseases **AND** any ADL/IADL restrictions | Subjective health: Poor **AND** >1 of chronic diseases | ALL | Subjective health: Poor | Numbers of chronic disease | Numbers of IADL restriction | EURO depression | Subjective health: Poor AND any ADL/IADL restrictions | >1 of chronic diseases **AND** any ADL/IADL restrictions | Subjective health: Poor **AND** >1 of chronic diseases | ALL |
| Total childhood deprivation score (Ref. Quintile 1) | | | | | | | | | | | | | | | | |
| Quintile 2 | 1.322*** | 1.224** | 1.357** | 1.099 | 1.343** | 1.368** | 1.226* | 1.327* | 1.131 | 1.406 | 1.500 | 1.585* | 1.499 | 1.424 | 1.346 | 1.430 |
| | (0.133) | (0.125) | (0.191) | (0.148) | (0.199) | (0.216) | (0.132) | (0.213) | (0.265) | (0.331) | (0.450) | (0.432) | (0.454) | (0.445) | (0.321) | (0.452) |
| Quintile 3 | 1.586*** | 1.332*** | 1.518*** | 1.039 | 1.522*** | 1.600*** | 1.327*** | 1.553*** | 1.356 | 1.620* | 1.303 | 1.795* | 1.210 | 1.305 | 1.705** | 1.257 |
| | (0.165) | (0.135) | (0.206) | (0.141) | (0.217) | (0.240) | (0.140) | (0.236) | (0.378) | (0.410) | (0.407) | (0.554) | (0.381) | (0.431) | (0.430) | (0.419) |
| Quintile 4 | 1.524*** | 1.389*** | 1.447*** | 1.060 | 1.432** | 1.537*** | 1.316** | 1.446** | 1.618* | 1.773** | 1.470 | 2.321*** | 1.471 | 1.545 | 1.741** | 1.601 |
| | (0.171) | (0.147) | (0.199) | (0.146) | (0.207) | (0.234) | (0.144) | (0.224) | (0.454) | (0.452) | (0.447) | (0.698) | (0.448) | (0.490) | (0.441) | (0.512) |
| Quintile 5 | 1.738*** | 1.761*** | 1.800*** | 1.697*** | 1.788*** | 2.050*** | 1.657*** | 1.896*** | 1.460 | 1.749** | 2.408*** | 2.759*** | 2.284*** | 2.552*** | 1.849** | 2.567*** |
| | (0.207) | (0.195) | (0.251) | (0.241) | (0.261) | (0.310) | (0.186) | (0.291) | (0.436) | (0.437) | (0.705) | (0.833) | (0.672) | (0.783) | (0.455) | (0.797) |
| Demographic indicators | | | | | | | | | | | | | | | | |
| Is a female | 0.883* | 1.216*** | 1.600*** | 1.664*** | 1.546*** | 1.699*** | 1.180** | 1.677*** | 1.578** | 1.818*** | 1.702*** | 2.098*** | 1.667** | 1.655** | 1.813*** | 1.660** |
| | (0.066) | (0.086) | (0.141) | (0.154) | (0.141) | (0.160) | (0.085) | (0.161) | (0.300) | (0.311) | (0.336) | (0.439) | (0.333) | (0.338) | (0.309) | (0.343) |
| Age | 1.017 | 1.041*** | 0.967* | 0.929*** | 0.969 | 0.985 | 1.037** | 0.982 | 1.106** | 1.121*** | 0.898*** | 0.885** | 0.904*** | 0.935 | 1.114*** | 0.934 |
| | (0.017) | (0.016) | (0.019) | (0.023) | (0.020) | (0.021) | (0.016) | (0.021) | (0.045) | (0.041) | (0.034) | (0.048) | (0.035) | (0.040) | (0.040) | (0.040) |
| Age squared | 1.089** | 0.995 | 1.259*** | 1.204*** | 1.242*** | 1.196*** | 1.013 | 1.198*** | 0.963 | 0.900 | 1.431*** | 1.295** | 1.404*** | 1.318*** | 0.918 | 1.319*** |
| | (0.040) | (0.031) | (0.050) | (0.056) | (0.051) | (0.050) | (0.033) | (0.051) | (0.094) | (0.072) | (0.113) | (0.133) | (0.114) | (0.116) | (0.071) | (0.117) |
| Education level (*Ref. None*) | | | | | | | | | | | | | | | | |
| Primary education | 2.209*** | 5.270 | 1.954 | 1.272* | 2.209*** | 5.270 | 1.954 | 1.272* | 0.980 | 0.961 | 0.159** | 2.264 | 0.139** | 0.181** | 0.762 | 0.161** |
| | (0.282) | (5.970) | (1.740) | (0.169) | (0.282) | (5.970) | (1.740) | (0.169) | (0.660) | (0.629) | (0.147) | (2.361) | (0.132) | (0.133) | (0.487) | (0.119) |
| Secondary education | 1.343*** | 4.098 | 1.540 | 1.114 | 1.343*** | 4.098 | 1.540 | 1.114 | 0.926 | 0.959 | 0.124** | 2.042 | 0.106** | 0.170** | 0.798 | 0.150*** |
| | (0.103) | (4.640) | (1.369) | (0.108) | (0.103) | (4.640) | (1.369) | (0.108) | (0.592) | (0.612) | (0.114) | (2.115) | (0.100) | (0.124) | (0.499) | (0.110) |
| Tertiary education | | 3.578 | 0.986 | | 0.869 | 2.174 | 2.619 | 1.979 | 0.736 | 0.903 | 0.097** | 2.404 | 0.090** | 0.150*** | 0.729 | 0.140*** |
| | | (4.057) | (0.879) | | (0.765) | (1.641) | (3.023) | (1.490) | (0.468) | (0.580) | (0.089) | (2.504) | (0.085) | (0.110) | (0.458) | (0.102) |
| **Country** (*Ref. Estonia*) | | | | | | | | | | | | | | | | |
| Latvia | 0.819** | 0.984 | 0.730*** | 0.695*** | 0.749*** | 0.796** | 1.002 | 0.814** | 0.591** | 0.751 | 0.544*** | 0.312*** | 0.536*** | 0.395*** | 0.715* | 0.392*** |
| | (0.064) | (0.073) | (0.068) | (0.069) | (0.071) | (0.078) | (0.076) | (0.081) | (0.125) | (0.148) | (0.120) | (0.094) | (0.119) | (0.098) | (0.140) | (0.097) |
| Employment status (*ref. Retired)* | | | | | | | | | | | | | | | | |
| Employed or self-employed | 0.648*** | 0.644*** | 0.554*** | 0.641*** | 0.456*** | 0.466*** | 0.597*** | 0.416*** | 0.842 | 0.916 | 0.411*** | 0.639 | 0.383*** | 0.396*** | 0.805 | 0.386*** |
| | (0.062) | (0.061) | (0.072) | (0.090) | (0.064) | (0.067) | (0.060) | (0.063) | (0.204) | (0.203) | (0.117) | (0.195) | (0.113) | (0.122) | (0.177) | (0.122) |
| Others | 2.558*** | 1.442*** | 2.199*** | 1.066 | 2.275*** | 2.217*** | 1.816*** | 2.204*** | 1.762* | 1.579 | 1.788* | 0.793 | 1.755* | 2.228** | 1.586 | 2.159** |
| | (0.406) | (0.188) | (0.322) | (0.201) | (0.338) | (0.349) | (0.240) | (0.348) | (0.541) | (0.440) | (0.553) | (0.344) | (0.544) | (0.723) | (0.445) | (0.707) |
| Household indicators | | | | | | | | | | | | | | | | |
| Is living with a partner | 0.911 | 0.996 | 0.866 | 0.929 | 0.874 | 0.828* | 0.968 | 0.826* | 1.068 | 1.221 | 1.081 | 0.690 | 1.140 | 1.093 | 1.259 | 1.170 |
| | (0.083) | (0.082) | (0.087) | (0.101) | (0.091) | (0.088) | (0.081) | (0.090) | (0.258) | (0.265) | (0.256) | (0.185) | (0.270) | (0.271) | (0.264) | (0.291) |
| Lives in Urban area | 0.974 | 1.109 | 0.973 | 1.010 | 0.974 | 1.019 | 1.062 | 0.997 | 1.215 | 2.244* | 2.492* | | 2.979** | 3.302* | 2.202* | 4.534* |
| | (0.074) | (0.078) | (0.080) | (0.088) | (0.082) | (0.091) | (0.076) | (0.090) | (0.441) | (1.076) | (1.224) | | (1.535) | (2.355) | (0.979) | (3.784) |
| Household size | 0.856** | 0.866** | 0.921 | 0.736*** | 0.904 | 0.872 | 0.864** | 0.849* | 0.938 | 1.113 | 0.677*** | 0.757 | 0.651*** | 0.736** | 1.106 | 0.704** |
| | (0.056) | (0.053) | (0.066) | (0.077) | (0.070) | (0.073) | (0.056) | (0.075) | (0.115) | (0.130) | (0.101) | (0.148) | (0.100) | (0.114) | (0.131) | (0.113) |
| Household income quartile (*ref. First quartile*) | | | | | | | | | | | | | | | | |
| Second quartile | 0.892 | 0.858 | 0.944 | 0.869 | 0.964 | 0.969 | 0.841* | 0.973 | 1.079 | 0.935 | 0.674 | 0.811 | 0.673 | 0.754 | 1.013 | 0.752 |
| | (0.095) | (0.081) | (0.105) | (0.134) | (0.110) | (0.116) | (0.080) | (0.119) | (0.270) | (0.209) | (0.163) | (0.282) | (0.164) | (0.192) | (0.223) | (0.194) |
| Third quartile | 0.681*** | 0.776** | 0.811 | 0.587*** | 0.776* | 0.785 | 0.695*** | 0.763* | 1.744** | 1.321 | 0.824 | 0.934 | 0.791 | 0.886 | 1.592* | 0.885 |
| | (0.082) | (0.086) | (0.113) | (0.100) | (0.113) | (0.122) | (0.080) | (0.122) | (0.477) | (0.341) | (0.256) | (0.368) | (0.252) | (0.297) | (0.411) | (0.302) |
| Fourth quartile | 0.555*** | 0.745** | 0.746* | 0.527*** | 0.755 | 0.699* | 0.684*** | 0.689** | 1.025 | 1.070 | 0.445** | 0.398** | 0.416** | 0.523 | 1.141 | 0.502* |
| | (0.092) | (0.106) | (0.125) | (0.120) | (0.130) | (0.128) | (0.100) | (0.129) | (0.357) | (0.337) | (0.164) | (0.186) | (0.157) | (0.208) | (0.356) | (0.204) |
| Difficulty in making ends meet: Yes | 1.877*** | 1.485*** | 2.275*** | 1.679*** | 2.281*** | 2.183*** | 1.576*** | 2.144*** | 1.561*** | 1.835*** | 2.003*** | 1.659** | 2.211*** | 2.088*** | 1.860*** | 2.262*** |
| | (0.121) | (0.088) | (0.166) | (0.137) | (0.173) | (0.172) | (0.095) | (0.172) | (0.262) | (0.286) | (0.342) | (0.338) | (0.378) | (0.366) | (0.285) | (0.401) |
| The respondent was born abroad | 1.023 | 1.080 | 0.907 | 1.207 | 0.950 | 1.038 | 1.073 | 1.070 | 1.010 | 1.132 | 0.837 | 1.045 | 0.856 | 0.934 | 1.032 | 0.934 |
| | (0.115) | (0.116) | (0.114) | (0.164) | (0.124) | (0.137) | (0.114) | (0.144) | (0.192) | (0.187) | (0.168) | (0.209) | (0.174) | (0.194) | (0.173) | (0.198) |
| Wave (ref. 7) | | | | | | | | | | | | | | | | |
| Wave 8 | 0.930 | 1.010 | 1.052 | | 1.080 | 1.103 | 1.004 | 1.138* | 0.985 | 0.923 | 1.239 | | 1.276 | 1.056 | 0.946 | 1.095 |
| | (0.057) | (0.054) | (0.069) | | (0.072) | (0.079) | (0.055) | (0.083) | (0.175) | (0.133) | (0.199) | | (0.207) | (0.180) | (0.139) | (0.189) |
| Wave 9 | 0.942 | 1.168*** | 0.755*** | 1.146** | 0.756*** | 0.874* | 1.074 | 0.864* | 0.715* | 1.070 | 0.749* | 1.170 | 0.794 | 0.817 | 0.994 | 0.868 |
| | (0.058) | (0.065) | (0.055) | (0.074) | (0.056) | (0.068) | (0.061) | (0.069) | (0.125) | (0.148) | (0.123) | (0.177) | (0.131) | (0.136) | (0.140) | (0.146) |
| Constant | 0.478 | 0.014*** | 0.374 | 37.919** | 0.363 | 0.041** | 0.018*** | 0.060* | 0.002** | 0.000*** | 443.389** | 380.596* | 283.153** | 12.705 | 0.000*** | 10.630 |
| | (0.482) | (0.020) | (0.573) | (57.713) | (0.567) | (0.064) | (0.027) | (0.094) | (0.006) | (0.000) | (1,125.521) | (1,309.202) | (739.945) | (35.796) | (0.000) | (30.816) |
| Observations | 7,216 | 7,221 | 7,221 | 3,990 | 7,221 | 7,221 | 7,221 | 7,221 | 1,373 | 1,373 | 1,373 | 715 | 1,373 | 1,373 | 1,373 | 1,373 |
| Log-likelihood | -3929 | -4544 | -3198 | -2282 | -3059 | -2821 | -4424 | -2751 | -596.2 | -779.4 | -676.7 | -444 | -668.5 | -644.9 | -793.9 | -637.2 |
| Pseudo R2 | 0.144 | 0.0917 | 0.182 | 0.0695 | 0.186 | 0.181 | 0.107 | 0.181 | 0.156 | 0.137 | 0.189 | 0.0876 | 0.190 | 0.191 | 0.146 | 0.194 |

Robust standard errors in parentheses (*** p<0.01, ** p<0.05, * p<0.1)

*Table A 5 Robustness check: Interaction between childhood deprivation and country/linguistic group (OR)*

| | Interaction between childhood deprivation and country | | | | | | | | Interaction between childhood deprivation and linguistic group | | | | | | | |
|---|---|---|---|---|---|---|---|---|---|---|---|---|---|---|---|---|
| VARIABLES | Subjective health: Poor | Numbers of chronic disease | Numbers of IADL restriction | EURO depression | Subjective health: Poor AND any ADL/IADL restrictions | >1 of chronic diseases **AND** any ADL/IADL restrictions | Subjective health: Poor **AND** >1 of chronic diseases | ALL | Subjective health: Poor | Numbers of chronic disease | Numbers of IADL restriction | EURO depression | Subjective health: Poor AND any ADL/IADL restrictions | >1 of chronic diseases **AND** any ADL/IADL restrictions | Subjective health: Poor **AND** >1 of chronic diseases | ALL |
| Total childhood deprivation score (Ref. Quintile 1) | | | | | | | | | | | | | | | | |
| Quintile 2 | 1.355*** | 1.374*** | 1.456*** | 1.108 | 1.451** | 1.487** | 1.302** | 1.458** | 1.029 | 1.376 | 1.447 | 1.300 | 1.434 | 1.327 | 1.284 | 1.323 |
| | (0.151) | (0.152) | (0.212) | (0.152) | (0.221) | (0.238) | (0.149) | (0.237) | (0.238) | (0.297) | (0.419) | (0.346) | (0.421) | (0.403) | (0.286) | (0.406) |
| Quintile 3 | 1.413*** | 1.338*** | 1.601*** | 1.067 | 1.586*** | 1.638*** | 1.283** | 1.597*** | 1.305 | 1.676** | 1.269 | 1.465 | 1.185 | 1.215 | 1.626** | 1.180 |
| | (0.158) | (0.144) | (0.221) | (0.146) | (0.230) | (0.251) | (0.142) | (0.248) | (0.328) | (0.383) | (0.352) | (0.414) | (0.333) | (0.359) | (0.371) | (0.352) |
| Quintile 4 | 1.467*** | 1.397*** | 1.570*** | 1.188 | 1.573*** | 1.582*** | 1.278** | 1.519*** | 1.934*** | 1.918*** | 1.424 | 1.791** | 1.444 | 1.412 | 1.951*** | 1.479 |
| | (0.177) | (0.157) | (0.219) | (0.163) | (0.230) | (0.243) | (0.146) | (0.237) | (0.481) | (0.409) | (0.374) | (0.462) | (0.381) | (0.389) | (0.428) | (0.410) |
| Quintile 5 | 1.638*** | 1.842*** | 1.973*** | 1.840*** | 1.984*** | 2.183*** | 1.727*** | 2.088*** | 1.808** | 2.026*** | 2.333*** | 2.077*** | 2.181*** | 2.194*** | 2.111*** | 2.208*** |
| | (0.211) | (0.216) | (0.276) | (0.260) | (0.289) | (0.329) | (0.204) | (0.320) | (0.463) | (0.427) | (0.556) | (0.527) | (0.527) | (0.559) | (0.453) | (0.569) |
| **Country** (*Ref. Estonia*) | | | | | | | | | | | | | | | | |
| Latvia | 0.786* | 1.100 | 0.940 | 0.575*** | 1.003 | 0.993 | 0.995 | 1.021 | 0.795*** | 0.964 | 0.704*** | 0.642*** | 0.720*** | 0.736*** | 0.963 | 0.747*** |
| | (0.113) | (0.168) | (0.207) | (0.117) | (0.228) | (0.243) | (0.156) | (0.252) | (0.059) | (0.067) | (0.059) | (0.061) | (0.062) | (0.066) | (0.068) | (0.068) |
| Israel | 0.155*** | 0.843 | 0.589** | 0.837 | 0.345*** | 0.673 | 0.294*** | 0.365*** | 0.183*** | 0.779*** | 0.720*** | 1.047 | 0.612*** | 0.681*** | 0.365*** | 0.606*** |
| | (0.027) | (0.134) | (0.145) | (0.210) | (0.103) | (0.183) | (0.055) | (0.118) | (0.017) | (0.066) | (0.078) | (0.129) | (0.069) | (0.078) | (0.033) | (0.072) |
| Language of interview (*Ref. National language*) | | | | | | | | | 0.571*** | 0.851 | 0.846 | 0.951 | 0.748 | 0.727 | 0.735* | 0.684 |
| | | | | | | | | | (0.097) | (0.140) | (0.190) | (0.201) | (0.173) | (0.177) | (0.128) | (0.170) |
| **Interactions** | | | | | | | | | | | | | | | | |

| | | | | | | | | |
|---|---|---|---|---|---|---|---|---|
| 1b.ch_totdepscore_q#1b.country | 1.000 | 1.000 | 1.000 | 1.000 | 1.000 | 1.000 | 1.000 | 1.000 |
| | (0.000) | (0.000) | (0.000) | (0.000) | (0.000) | (0.000) | (0.000) | (0.000) |
| 1b.ch_totdepscore_q#3o.country | 1.000 | 1.000 | 1.000 | 1.000 | 1.000 | 1.000 | 1.000 | 1.000 |
| | (0.000) | (0.000) | (0.000) | (0.000) | (0.000) | (0.000) | (0.000) | (0.000) |
| 1b.ch_totdepscore_q#4o.country | 1.000 | 1.000 | 1.000 | 1.000 | 1.000 | 1.000 | 1.000 | 1.000 |
| | (0.000) | (0.000) | (0.000) | (0.000) | (0.000) | (0.000) | (0.000) | (0.000) |
| 2o.ch_totdepscore_q#1b.country | 1.000 | 1.000 | 1.000 | 1.000 | 1.000 | 1.000 | 1.000 | 1.000 |
| | (0.000) | (0.000) | (0.000) | (0.000) | (0.000) | (0.000) | (0.000) | (0.000) |
| 2.ch_totdepscore_q#3.country | 0.778 | 0.665* | 0.743 | 1.282 | 0.719 | 0.627 | 0.770 | 0.620 |
| | (0.157) | (0.139) | (0.214) | (0.364) | (0.212) | (0.203) | (0.165) | (0.202) |
| 2.ch_totdepscore_q#4.country | 0.610* | 0.920 | 0.567 | 1.213 | 0.672 | 0.534 | 0.672 | 0.736 |
| | (0.170) | (0.218) | (0.204) | (0.440) | (0.284) | (0.207) | (0.198) | (0.327) |
| 3o.ch_totdepscore_q#1b.country | 1.000 | 1.000 | 1.000 | 1.000 | 1.000 | 1.000 | 1.000 | 1.000 |
| | (0.000) | (0.000) | (0.000) | (0.000) | (0.000) | (0.000) | (0.000) | (0.000) |
| 3.ch_totdepscore_q#3.country | 1.251 | 0.980 | 0.683 | 1.335 | 0.676 | 0.659 | 1.139 | 0.661 |
| | (0.271) | (0.208) | (0.190) | (0.384) | (0.193) | (0.202) | (0.246) | (0.203) |
| 3.ch_totdepscore_q#4.country | 1.004 | 0.762 | 0.712 | 1.096 | 1.155 | 0.579 | 1.052 | 1.031 |
| | (0.242) | (0.170) | (0.240) | (0.392) | (0.441) | (0.209) | (0.269) | (0.419) |
| 4o.ch_totdepscore_q#1b.country | 1.000 | 1.000 | 1.000 | 1.000 | 1.000 | 1.000 | 1.000 | 1.000 |
| | (0.000) | (0.000) | (0.000) | (0.000) | (0.000) | (0.000) | (0.000) | (0.000) |
| 4.ch_totdepscore_q#3.country | 1.004 | 0.996 | 0.655 | 1.017 | 0.618* | 0.749 | 1.134 | 0.756 |
| | (0.221) | (0.213) | (0.180) | (0.287) | (0.175) | (0.225) | (0.247) | (0.229) |
| 4.ch_totdepscore_q#4.country | 1.363 | 0.899 | 1.409 | 1.734 | 2.052** | 1.249 | 1.570* | 2.076* |
| | (0.342) | (0.197) | (0.442) | (0.596) | (0.748) | (0.424) | (0.395) | (0.807) |
| 5o.ch_totdepscore_q#1b.country | 1.000 | 1.000 | 1.000 | 1.000 | 1.000 | 1.000 | 1.000 | 1.000 |
| | (0.000) | (0.000) | (0.000) | (0.000) | (0.000) | (0.000) | (0.000) | (0.000) |
| 5.ch_totdepscore_q#3.country | 0.957 | 0.726 | 0.772 | 0.930 | 0.700 | 0.734 | 0.736 | 0.691 |
| | (0.219) | (0.152) | (0.210) | (0.263) | (0.193) | (0.215) | (0.154) | (0.203) |

| | | | | | | | | | | | | | | | | |
|---|---|---|---|---|---|---|---|---|---|---|---|---|---|---|---|---|
| 5.ch_totdepscore_q#4.country | 1.838** | 0.760 | 2.203*** | 1.061 | 3.199*** | 1.332 | 1.456 | 2.457** | | | | | | | | |
| | (0.459) | (0.177) | (0.670) | (0.360) | (1.106) | (0.432) | (0.364) | (0.910) | | | | | | | | |
| 1b.ch_totdepscore_q#0b.lingua | | | | | | | | | 1.000 | 1.000 | 1.000 | 1.000 | 1.000 | 1.000 | 1.000 | 1.000 |
| | | | | | | | | | (0.000) | (0.000) | (0.000) | (0.000) | (0.000) | (0.000) | (0.000) | (0.000) |
| 1b.ch_totdepscore_q#1o.lingua | | | | | | | | | 1.000 | 1.000 | 1.000 | 1.000 | 1.000 | 1.000 | 1.000 | 1.000 |
| | | | | | | | | | (0.000) | (0.000) | (0.000) | (0.000) | (0.000) | (0.000) | (0.000) | (0.000) |
| 2o.ch_totdepscore_q#0b.lingua | | | | | | | | | 1.000 | 1.000 | 1.000 | 1.000 | 1.000 | 1.000 | 1.000 | 1.000 |
| | | | | | | | | | (0.000) | (0.000) | (0.000) | (0.000) | (0.000) | (0.000) | (0.000) | (0.000) |
| 2.ch_totdepscore_q#1.lingua | | | | | | | | | 1.214 | 0.888 | 0.904 | 0.897 | 0.934 | 0.977 | 0.918 | 1.006 |
| | | | | | | | | | (0.300) | (0.208) | (0.287) | (0.261) | (0.302) | (0.328) | (0.223) | (0.342) |
| 3o.ch_totdepscore_q#0b.lingua | | | | | | | | | 1.000 | 1.000 | 1.000 | 1.000 | 1.000 | 1.000 | 1.000 | 1.000 |
| | | | | | | | | | (0.000) | (0.000) | (0.000) | (0.000) | (0.000) | (0.000) | (0.000) | (0.000) |
| 3.ch_totdepscore_q#1.lingua | | | | | | | | | 1.179 | 0.741 | 1.176 | 0.742 | 1.335 | 1.244 | 0.803 | 1.348 |
| | | | | | | | | | (0.315) | (0.181) | (0.354) | (0.227) | (0.412) | (0.401) | (0.197) | (0.442) |
| 4o.ch_totdepscore_q#0b.lingua | | | | | | | | | 1.000 | 1.000 | 1.000 | 1.000 | 1.000 | 1.000 | 1.000 | 1.000 |
| | | | | | | | | | (0.000) | (0.000) | (0.000) | (0.000) | (0.000) | (0.000) | (0.000) | (0.000) |
| 4.ch_totdepscore_q#1.lingua | | | | | | | | | 0.777 | 0.676* | 1.093 | 0.655 | 1.103 | 1.132 | 0.669* | 1.085 |
| | | | | | | | | | (0.206) | (0.155) | (0.314) | (0.184) | (0.322) | (0.344) | (0.159) | (0.334) |
| 5o.ch_totdepscore_q#0b.lingua | | | | | | | | | 1.000 | 1.000 | 1.000 | 1.000 | 1.000 | 1.000 | 1.000 | 1.000 |
| | | | | | | | | | (0.000) | (0.000) | (0.000) | (0.000) | (0.000) | (0.000) | (0.000) | (0.000) |
| 5.ch_totdepscore_q#1.lingua | | | | | | | | | 1.012 | 0.786 | 0.877 | 0.855 | 0.970 | 0.975 | 0.771 | 0.976 |
| | | | | | | | | | (0.275) | (0.177) | (0.232) | (0.237) | (0.263) | (0.276) | (0.179) | (0.282) |
| Demographic indicators | | | | | | | | | | | | | | | | |
| Is a female | 0.952 | 1.188*** | 1.515*** | 1.738*** | 1.488*** | 1.591*** | 1.217*** | 1.589*** | 0.957 | 1.195*** | 1.497*** | 1.746*** | 1.472*** | 1.580*** | 1.220*** | 1.578*** |
| | (0.061) | (0.071) | (0.112) | (0.138) | (0.114) | (0.125) | (0.075) | (0.128) | (0.061) | (0.071) | (0.111) | (0.139) | (0.113) | (0.124) | (0.075) | (0.128) |
| Age | 1.019 | 1.061*** | 0.953*** | 0.926*** | 0.955*** | 0.979 | 1.047*** | 0.975 | 1.022 | 1.064*** | 0.952*** | 0.927*** | 0.954*** | 0.981 | 1.051*** | 0.976 |
| | (0.014) | (0.014) | (0.016) | (0.019) | (0.016) | (0.017) | (0.014) | (0.018) | (0.014) | (0.014) | (0.016) | (0.019) | (0.016) | (0.018) | (0.014) | (0.018) |
| Age squared | 1.080** | 0.969 | 1.295*** | 1.216*** | 1.276*** | 1.212*** | 0.998 | 1.212*** | 1.073** | 0.963 | 1.295*** | 1.214*** | 1.277*** | 1.208*** | 0.992 | 1.210*** |
| | (0.034) | (0.027) | (0.043) | (0.049) | (0.044) | (0.043) | (0.028) | (0.044) | (0.034) | (0.027) | (0.043) | (0.049) | (0.044) | (0.043) | (0.028) | (0.044) |
| Education level (*Ref. None*) | | | | | | | | | | | | | | | | |
| Primary education | 0.452 | 1.356 | 0.993 | 1.154 | 0.955 | 1.207 | 0.935 | 1.237 | 0.359 | 1.358 | 0.824 | 0.896 | 0.775 | 1.024 | 0.772 | 1.008 |
| | (0.303) | (0.573) | (0.389) | (0.718) | (0.378) | (0.482) | (0.401) | (0.489) | (0.229) | (0.572) | (0.321) | (0.579) | (0.307) | (0.395) | (0.343) | (0.387) |
| Secondary education | 0.287* | 1.094 | 0.769 | 0.982 | 0.706 | 0.930 | 0.697 | 0.925 | 0.217** | 1.086 | 0.634 | 0.754 | 0.567 | 0.783 | 0.562 | 0.746 |
| | (0.192) | (0.460) | (0.300) | (0.609) | (0.278) | (0.370) | (0.298) | (0.364) | (0.138) | (0.456) | (0.246) | (0.487) | (0.224) | (0.301) | (0.249) | (0.285) |
| Tertiary education | 0.213** | 0.994 | 0.543 | 0.959 | 0.496* | 0.673 | 0.581 | 0.663 | 0.162*** | 0.983 | 0.456** | 0.736 | 0.406** | 0.573 | 0.469* | 0.541 |
| | (0.142) | (0.419) | (0.213) | (0.596) | (0.197) | (0.269) | (0.249) | (0.262) | (0.103) | (0.413) | (0.178) | (0.476) | (0.161) | (0.222) | (0.208) | (0.209) |
| Employment status (*ref. Retired)* | | | | | | | | | | | | | | | | |
| Employed or self-employed | 0.611*** | 0.687*** | 0.517*** | 0.684*** | 0.423*** | 0.451*** | 0.605*** | 0.397*** | 0.620*** | 0.688*** | 0.522*** | 0.681*** | 0.428*** | 0.454*** | 0.612*** | 0.402*** |
| | (0.049) | (0.052) | (0.057) | (0.078) | (0.050) | (0.055) | (0.050) | (0.051) | (0.049) | (0.052) | (0.057) | (0.077) | (0.050) | (0.055) | (0.050) | (0.052) |
| Others | 1.988*** | 1.576*** | 2.282*** | 1.187 | 2.200*** | 2.454*** | 1.762*** | 2.293*** | 2.009*** | 1.560*** | 2.332*** | 1.187 | 2.258*** | 2.495*** | 1.793*** | 2.354*** |
| | (0.219) | (0.152) | (0.252) | (0.168) | (0.249) | (0.288) | (0.175) | (0.274) | (0.222) | (0.151) | (0.254) | (0.167) | (0.251) | (0.290) | (0.179) | (0.278) |
| Household indicators | | | | | | | | | | | | | | | | |
| Is living with a partner | 0.858* | 0.955 | 0.864* | 0.873 | 0.888 | 0.865 | 0.959 | 0.885 | 0.865* | 0.960 | 0.862* | 0.875 | 0.886 | 0.867 | 0.966 | 0.889 |
| | (0.069) | (0.069) | (0.076) | (0.082) | (0.081) | (0.080) | (0.071) | (0.084) | (0.070) | (0.069) | (0.075) | (0.082) | (0.081) | (0.081) | (0.071) | (0.085) |
| Lives in Urban area | 1.012 | 1.157** | 1.007 | 1.072 | 1.018 | 1.086 | 1.132* | 1.072 | 0.954 | 1.103 | 0.978 | 1.032 | 0.978 | 1.047 | 1.064 | 1.026 |
| | (0.073) | (0.079) | (0.079) | (0.090) | (0.082) | (0.092) | (0.079) | (0.092) | (0.069) | (0.075) | (0.078) | (0.088) | (0.080) | (0.091) | (0.075) | (0.090) |
| Household size | 0.863*** | 0.875*** | 0.849*** | 0.731*** | 0.838*** | 0.810*** | 0.876** | 0.795*** | 0.872*** | 0.877** | 0.846*** | 0.735*** | 0.836*** | 0.810*** | 0.883** | 0.793*** |
| | (0.046) | (0.045) | (0.053) | (0.063) | (0.055) | (0.057) | (0.048) | (0.059) | (0.046) | (0.045) | (0.053) | (0.062) | (0.055) | (0.058) | (0.048) | (0.059) |
| Household income quartile (*ref. First quartile*) | | | | | | | | | | | | | | | | |
| Second quartile | 0.867 | 0.853* | 0.875 | 0.861 | 0.883 | 0.871 | 0.841** | 0.879 | 0.892 | 0.866* | 0.874 | 0.863 | 0.883 | 0.873 | 0.860* | 0.883 |
| | (0.079) | (0.069) | (0.083) | (0.115) | (0.086) | (0.089) | (0.070) | (0.092) | (0.082) | (0.071) | (0.083) | (0.115) | (0.087) | (0.090) | (0.072) | (0.093) |
| Third quartile | 0.727*** | 0.795** | 0.798* | 0.627*** | 0.787* | 0.777* | 0.739*** | 0.775* | 0.752*** | 0.810** | 0.783** | 0.633*** | 0.774** | 0.772** | 0.755*** | 0.768* |
| | (0.074) | (0.074) | (0.094) | (0.093) | (0.097) | (0.101) | (0.073) | (0.104) | (0.076) | (0.075) | (0.092) | (0.093) | (0.096) | (0.101) | (0.074) | (0.104) |
| Fourth quartile | 0.551*** | 0.718*** | 0.649*** | 0.481*** | 0.660*** | 0.618*** | 0.674*** | 0.616*** | 0.565*** | 0.730*** | 0.638*** | 0.484*** | 0.650*** | 0.617*** | 0.688*** | 0.615*** |
| | (0.076) | (0.087) | (0.093) | (0.093) | (0.097) | (0.096) | (0.085) | (0.098) | (0.078) | (0.089) | (0.091) | (0.092) | (0.096) | (0.096) | (0.087) | (0.098) |
| Difficulty in making ends meet: Yes | 1.857*** | 1.527*** | 2.227*** | 1.697*** | 2.247*** | 2.148*** | 1.670*** | 2.143*** | 1.784*** | 1.479*** | 2.194*** | 1.644*** | 2.198*** | 2.099*** | 1.602*** | 2.083*** |
| | (0.102) | (0.079) | (0.140) | (0.120) | (0.146) | (0.144) | (0.087) | (0.146) | (0.099) | (0.076) | (0.138) | (0.118) | (0.142) | (0.140) | (0.084) | (0.142) |
| The respondent was born abroad | 1.385*** | 1.403*** | 1.019 | 1.298*** | 1.109 | 1.226** | 1.457*** | 1.289*** | 1.158* | 1.210** | 0.963 | 1.165 | 1.016 | 1.095 | 1.190** | 1.123 |
| | (0.106) | (0.096) | (0.080) | (0.113) | (0.089) | (0.101) | (0.101) | (0.107) | (0.096) | (0.092) | (0.087) | (0.115) | (0.095) | (0.104) | (0.092) | (0.109) |
| Wave (ref. 7) | | | | | | | | | | | | | | | | |
| Wave 8 | 1.002 | 1.071 | 1.123** | | 1.153** | 1.155** | 1.069 | 1.205*** | 1.012 | 1.073 | 1.138** | | 1.168*** | 1.166** | 1.080 | 1.219*** |
| | (0.053) | (0.050) | (0.064) | | (0.067) | (0.072) | (0.051) | (0.076) | (0.053) | (0.050) | (0.064) | | (0.068) | (0.072) | (0.052) | (0.076) |
| Wave 9 | 0.955 | 1.254*** | 0.764*** | 1.123** | 0.778*** | 0.897 | 1.128** | 0.904 | 0.945 | 1.240*** | 0.769*** | 1.109* | 0.781*** | 0.896 | 1.118** | 0.904 |
| | (0.052) | (0.061) | (0.049) | (0.062) | (0.051) | (0.060) | (0.057) | (0.062) | (0.051) | (0.060) | (0.049) | (0.062) | (0.051) | (0.060) | (0.056) | (0.062) |
| Constant | 1.980 | 0.014*** | 2.095 | 47.021*** | 1.980 | 0.232 | 0.038*** | 0.314 | 3.542 | 0.015*** | 3.325 | 60.291*** | 3.639 | 0.372 | 0.052*** | 0.552 |
| | (2.147) | (0.013) | (2.332) | (67.270) | (2.282) | (0.278) | (0.036) | (0.380) | (3.794) | (0.014) | (3.709) | (87.543) | (4.201) | (0.446) | (0.050) | (0.673) |
| Observations | 10,079 | 10,079 | 10,079 | 5,318 | 10,079 | 10,079 | 10,079 | 10,079 | 10,079 | 10,079 | 10,079 | 5,318 | 10,079 | 10,079 | 10,079 | 10,079 |
| Log-likelihood | -5340 | -6248 | -4401 | -3077 | -4202 | -3936 | -5953 | -3825 | -5330 | -6240 | -4416 | -3076 | -4217 | -3941 | -5940 | -3830 |
| Pseudo R2 | 0.190 | 0.104 | 0.195 | 0.0784 | 0.201 | 0.195 | 0.137 | 0.196 | 0.192 | 0.106 | 0.193 | 0.0787 | 0.198 | 0.194 | 0.139 | 0.195 |

Robust standard errors in parentheses (*** p<0.01, ** p<0.05, * p<0.1)